\documentclass[aps,pra,showpacs,superscriptaddress,amssymb,twocolumn]{revtex4-1}
\usepackage{graphicx}
\usepackage{appendix}
\usepackage{color}
\usepackage[caption=false]{subfig}
\usepackage{epstopdf} 
\DeclareGraphicsExtensions{.pdf,.eps,.png,.jpg,.mps}
\usepackage{hyperref}
\usepackage{bm,amsmath}
\usepackage{dcolumn}

\newcommand{\ket}[1]{\left|#1\right\rangle}

\def\BEq{\begin{equation}}
\def\EEq{\end{equation}}
\def\BEqA{\begin{eqnarray}}
\def\EEqA{\end{eqnarray}}
\def\BW{\begin{widetext}}
\def\EW{\end{widetext}}

\begin{document}

\title{Surface code with decoherence: An analysis of three superconducting architectures}

\author{Joydip Ghosh}
\email{joydip.ghosh@gmail.com}
\affiliation{Department of Physics and Astronomy, University of Georgia, Athens, Georgia 30602, USA}
\author{Austin G. Fowler}
\email{austingfowler@gmail.com}
\affiliation{Centre for Quantum Computation and Communication Technology, School of Physics, The University of Melbourne, Victoria 3010, Australia}
\author{Michael R. Geller}
\email{mgeller@uga.edu}
\affiliation{Department of Physics and Astronomy, University of Georgia, Athens, Georgia 30602, USA}

\date{\today}

\begin{abstract}
We consider realistic, multi-parameter error models and investigate the performance of the surface code for three possible fault-tolerant superconducting quantum computer architectures. We map amplitude and phase damping to a diagonal Pauli ``depolarization" channel via the Pauli twirl approximation, and obtain the logical error rate as a function of the qubit $T_{1,2}$ and state preparation, gate, and readout errors. A numerical Monte Carlo simulation is performed to obtain the logical error rates, and a leading-order analytic formula is derived to estimate their behavior below threshold. Our results suggest that scalable fault-tolerant quantum computation should be possible with existing superconducting devices.
\end{abstract}


\maketitle

\section{Introduction}
\label{sec:Introduction}

\begin{figure}[htb]
\includegraphics[angle=0,width=0.88\linewidth]{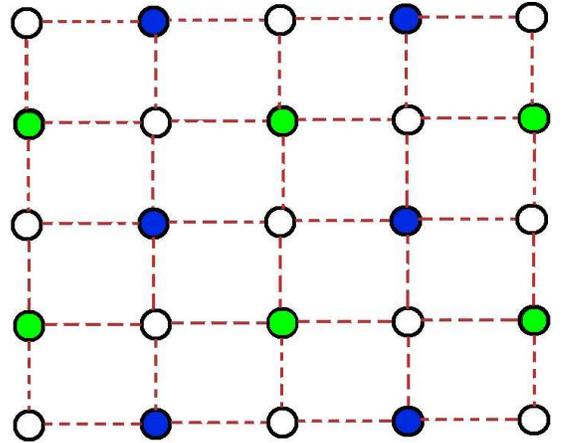}
\caption{(Color online) Layout of the distance-3 surface code considered here. Open circles denote data qubits, and light green (dark blue) filled circles denote $X$-type ($Z$-type) syndrome qubits. The dashed lines denote tunable qubit-qubit coupling. We refer to this hardware design as the textbook architecture.}
\label{fig:textbook}
\end{figure}

The surface code is a topological stabilizer code that is attractive because of its two-dimensional nearest-neighbor layout and high fault-tolerant error threshold \cite{1998quant.ph.11052B,dennis:4452,PhysRevLett.98.190504,1367-2630-9-6-199,Wang11,Fowl11b,PhysRevA.86.032324}. The most direct implementation of the surface code with superconducting circuits leads to the hardware design shown in Fig.~\ref{fig:textbook}, where the circles represent qubit devices and the dotted lines between them represent tunable couplers. These tunable couplers, however, significantly increase the complexity of the hardware. We therefore analyze the performance of this most basic design, which we call the {\it textbook} architecture, as well as two other fault-tolerant architectures for superconducting qubits with fixed capacitive coupling, using a (mostly) realistic error model that includes qubit decoherence. Although our approach is valid for large qubit arrays, we especially focus on first-generation implementations with code distances $d=3$ and $d=5$, and show that an experimental demonstration of a small-$d$ topological quantum memory should be possible with existing superconducting devices; the $d=5$ case already exhibits a pronounced quantum memory enhancement with current transmon $T_1$ values.

\begin{figure}[htb]
\includegraphics[angle=0,width=0.9\linewidth]{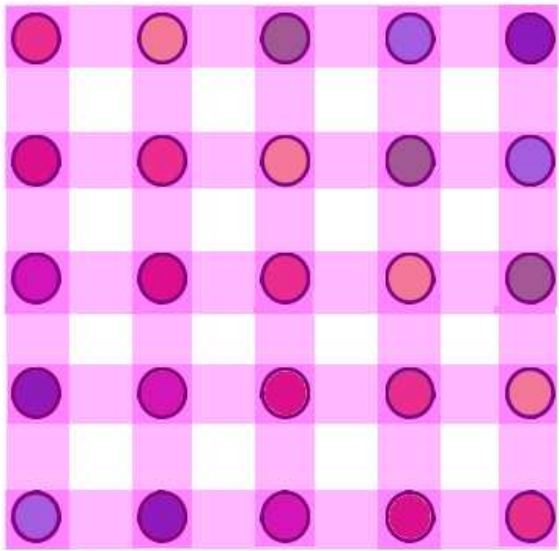}
\caption{(Color online) Schematic diagram of the distance-3 Helmer architecture. The circles represent superconducting qubits, with ``idle" frequencies indicated by their colors. The horizontal and vertical magenta (gray) rectangles are resonators. All horizontal (vertical) resonators have the same frequency.}
\label{fig:Helmer}
\end{figure}

\begin{figure}[htb]
\includegraphics[angle=-1,width=\linewidth]{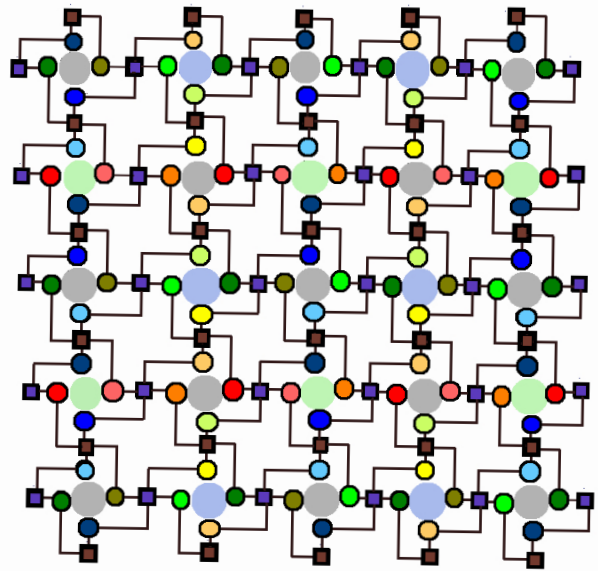}
\caption{(Color online) Schematic diagram of the architecture discussed by DiVincenzo \cite{1402-4896-2009-T137-014020} for code distance $d \! = \! 3$. The filled circles with boundaries represent qubits, squares with boundaries represent resonators, and colors of both denote their fixed frequencies. The unbounded circles are for the eye and indicate whether a given block is for data (dark gray), $X$-type syndrome (light green), or $Z$-type syndrome (blue). A possible frequency allocation for all the components is shown.}
\label{fig:IBM}
\end{figure}

Figures \ref{fig:textbook} through \ref{fig:IBM} show diagrams of the three fault-tolerant superconducting architectures we consider. For the textbook architecture we assume a two-dimensional array of transmon qubits \cite{PhysRevA.76.042319,PhysRevB.77.180502,PhysRevLett.107.240501,PhysRevB.86.100506}, and tunable couplers \cite{Hime01122006,PhysRevLett.98.057004,Niskanen04052007,PhysRevLett.98.177001,PhysRevLett.104.177004,PhysRevB.82.104522,PhysRevLett.106.060501,PhysRevB.84.144516} connecting nearest neighbors. The transmon qubits have tunable frequency \cite{DiCarlo2009}. We assume that the CNOT gates used to implement the stabilizer measurements are performed using the CZ gate of Strauch {\it et al.} \cite{StrauchPRL03}, which has extremely high performance in realistic, multi-qubit settings \cite{ghosh2012CZ}. High-fidelity single-qubit gates are carried out using DRAG pulses \cite{motzoi2009prl}. The initial states of the syndrome qubits are prepared via ideal projective measurements followed by local rotations (if required). We assume the ``catch-disperse-release" measurement protocol of Ref.~\footnote{E. A. Sete, A. Galiautdinov, E. Mlinar, J. M. Martinis, and A. N. Korotkov, unpublished.} for this purpose, as well as for syndrome readout. Ideal tunable coupler performance---including infinite on/off ratio---is assumed, and stray coupling (such as that arising from unintended capacitance between device elements) is ignored. Gate parameters adopted for this architecture are discussed in Sec.~\ref{sec:architectures} and summarized in Table \ref{table:model}. 

Although tunable couplers have been demonstrated by several groups, it is unknown whether they will be practical for use in a large-scale quantum computer. Therefore we consider two alternative architectures with fixed (capacitive) coupling. Figure \ref{fig:Helmer} shows an architecture proposed by Helmer \textit{et al.} \cite{0295-5075-85-5-50007}, where each qubit in a two-dimensional square lattice is coupled to a ``horizontal" as well as a ``vertical" cavity. Horizontal and vertical cavities in this architecture are fixed at different frequencies, while qubit frequencies are varied between them. The CNOT gates between a pair of adjacent qubits across the cavity are performed via an effective two-qubit flip-flop interaction in the dispersive regime \cite{0295-5075-85-5-50007}. While parallel CNOT gates between two or more pairs of qubits attached to the same resonator are allowed \cite{0295-5075-85-5-50007}, these simultaneous operations reduce the gate fidelity. Note that the Helmer architecture is not scalable, but the small distance cases are still of interest here.

Finally, we consider a scalable fixed-coupling architecture discussed by DiVincenzo \cite{1402-4896-2009-T137-014020} and shown schematically in Fig.~\ref{fig:IBM}.  Here each qubit (circle with solid boundary) is coupled to two resonators (squares with solid boundary). Both the qubit and resonator frequencies are fixed, avoiding the need for low-frequency qubit biases, and the CNOT gates are performed via a cross-resonance protocol using microwaves \cite{PhysRevLett.107.080502,2012arXiv1202.5344C,PhysRevB.81.134507}. Note that the number of qubit frequencies required for this architecture is independent of the number of qubits in a large array. In Fig.~\ref{fig:IBM} we have shown a possible frequency allocation represented by 14 colors, 12 for qubits and 2 for resonators. We emphasize that the error model considered here for the DiVincenzo architecture ignores the effects of higher-order interactions, microwave cross-coupling, and other multi-qubit errors typically neglected in theoretical models, which might be significant in this architecture given the larger values of coupling required.

We analyze these different architectures by fixing the intrinsic errors and gate times to estimated realistic values and calculating the logical error rate as a function of the qubit coherence time $T_1$. For tunable transmon qubits, the  $T_2$ time is assumed to be equal to $T_1$, while for fixed-frequency transmons we assume that $T_2=2T_1$. The logical error rate is calculated by mapping amplitude and phase damping to the asymmetric ``depolarization" channel (ADC), a single-qubit error channel that is diagonal in the Pauli basis. This is explained in Sec.~\ref{sec:QuantumNoise}. The depolarization channel error model is widely used in the quantum error correction literature, and the symmetric case allows simple comparison (especially of fault-tolerant error-threshold values) between different error-correcting codes. The action of the depolarization channel on stabilizer states can be efficiently simulated with a classical computer, enabling the direct calculation of logical error rates for large distance codes, and it accurately captures pure dephasing (but only approximately describes the decoherence found in real superconducting qubits). In Sec.~\ref{sec:SurfaceCode} we derive a leading-order analytic expression for the logical error rate that estimates the below-threshold scaling behavior (for small code distances). Section \ref{sec:architectures} gives the approximate performance of the three fault-tolerant architectures discussed above, using both the leading-order analytic formula and classical Monte Carlo simulation. There are many ways to implement a surface code with superconducting qubits, and the design details of any given fault-tolerant architecture will surely be improved and optimized over time; in this sense the architectures of Figs.~\ref{fig:Helmer} and \ref{fig:IBM} mainly serve as examples of our approach and indicate that large-scale quantum computers should be possible with existing superconducting devices (assuming the simple error models considered here). The textbook architecture of Fig.~\ref{fig:textbook} is also interesting because it likely provides a {\it bound} on the performance of any possible future superconducting surface code implementation, as the additional error-correction-cycle steps and unwanted multi-qubit interactions of alternative fixed-coupling designs will only degrade the performance. Comparing the performance of the textbook architecture with the
Helmer and DiVincenzo architectures also allows one to assess the benefit of using tunable couplers, which increases the hardware complexity but offers a lower $T_1$ threshold and logical error rate.

\section{MAPPING DECOHERENCE TO A DIAGONAL PAULI CHANNEL}
\label{sec:QuantumNoise}

In this section we discuss the use of Pauli twirling \cite{PhysRevA.72.052326,Emerson28092007,PhysRevA.78.012347,PhysRevA.80.012304,Sarvepalli08052009} to approximately model qubit decoherence by an asymmetric depolarization channel, which---by the Gottesman-Knill theorem---makes efficient classical Monte Carlo simulation possible.

\subsection{Amplitude and phase damping}

Quantum systems coupled to an environment undergo spontaneous dissipation of energy, which is usually modeled by the amplitude damping channel. For a single qubit this has the form 
\BEq
\rho \rightarrow
\mathcal{E}_{\rm AD}\left(\rho\right)=E^{\rm AD}_{1}{\rho} \, E_{1}^{{\rm AD} \dagger}+E^{\rm AD}_{2}{\rho} \, E_{2}^{{\rm AD}\dagger},
\EEq
where
\BEq
E^{\rm AD}_{1}=\begin{pmatrix}
 1 & 0 \\
 0 & \sqrt{1-p_{\rm AD}} \\
 \end{pmatrix} \  {\rm and} \ \
E^{\rm AD}_{2}=\begin{pmatrix}
 0 & \sqrt{p_{\rm AD}} \\
 0 & 0 \\
 \end{pmatrix}. 
\EEq
The $E^{\rm AD}_{m}$ are Kraus matrices for the amplitude damping channel, and $p_{\rm AD}$ can be interpreted as the probability of a single photon emission from the qubit. 

Phase damping or pure dephasing is a decoherence process generated by random phase kicks on a single qubit. Assuming the phase kick angle is a Gaussian-distributed random variable, the Kraus matrices for this process are
\BEq
E^{\rm PD}_{1}=\begin{pmatrix}
 1 & 0 \\
 0 & \sqrt{1-p_{\rm PD}} \\
 \end{pmatrix}  \  {\rm and} \ \
E^{\rm PD}_{2}=\begin{pmatrix}
 0 & 0 \\
 0 & \sqrt{p_{\rm PD}} \\
 \end{pmatrix}. 
\EEq

The combined channel of amplitude and phase damping can also be described by a set of three Kraus matrices,
\begin{eqnarray}
E^{\rm D}_{1} &=& \begin{pmatrix}
 1 & 0 \\
 0 & \sqrt{1-\gamma - \lambda} \\
 \end{pmatrix} \nonumber \\ &=& \frac{1+\sqrt{1-\gamma - \lambda}}{2}\mathbb{I} + \frac{1-\sqrt{1-\gamma - \lambda}}{2}\sigma^{\rm z}, \nonumber \\ 
 E^{\rm D}_{2} &=& \begin{pmatrix}
 0 & \sqrt{\gamma} \\
 0 & 0 \\
 \end{pmatrix} = \frac{\sqrt{\gamma}}{2}\sigma^{\rm x}+\frac{i\sqrt{\gamma}}{2}\sigma^{\rm y}, \nonumber \\
E^{\rm D}_{3} &=& \begin{pmatrix}
 0 & 0 \\
 0 & \sqrt{\lambda} \\
 \end{pmatrix}=\frac{\sqrt{\lambda}}{2}\mathbb{I}-\frac{\sqrt{\lambda}}{2}\sigma^{\rm z},
\label{eq:xdef}
\end{eqnarray}
where, $\gamma \equiv p_{\rm AD}$ and $\lambda \equiv (1-p_{\rm AD})p_{\rm PD}$. Next we represent the parameters $p_{\rm AD}$ and $p_{\rm PD}$ in terms of the single-qubit relaxation time $T_{1}$ and dephasing time $T_{2}$, 
\begin{eqnarray}
&1-p_{\rm AD} = e^{-t/T_{\rm 1}}, & \\
&\sqrt{\left(1-p_{\rm AD}\right)\left(1-p_{\rm PD}\right)} =e^{-t/T_{\rm 2}}. &
\end{eqnarray}
The combination of amplitude and phase damping on a single qubit transforms the density matrix as,
\BEq
\rho \rightarrow 
\mathcal{E}_{\rm D}\left(\rho\right)=\begin{pmatrix}
 1-\rho_{11}e^{-t/T_{1}} & \rho_{01} \, e^{-t/T_{2}} \\
 \rho_{01}^* \, e^{-t/T_{2}} & \rho_{11}e^{-t/T_{1}} \\
 \end{pmatrix}.
 \label{qubit with decoherence}
\EEq

\subsection{Asymmetric depolarization channel}

Classical simulation of Eq.~(\ref{qubit with decoherence}) is inefficient for a multi-qubit system. For example, the textbook architecture requires $25$ physical qubits for $d=3$ and $81$ physical qubits for $d\!=\!5$. The dimension of the Hilbert space is more than $33$ million for $d\!=\!3$ and more than $10^{24}$ for $d\!=\!5$. This motivates one to construct a simplified error model which is tractable via some efficient classical simulation.

The asymmetric depolarization channel (ADC) is such a model, where a decoherent qubit is assumed to suffer from discrete Pauli $X$ (bit-flip) errors, $Z$ (phase flip) errors, or $Y$ (both):
\BEq
\label{eq:depolarizationChannel}
\mathcal{E}_{\rm ADC}\left(\rho\right)=(1-p_\Sigma){\rho}+p_{X}X{\rho}X+p_{Y}Y{\rho}Y+p_{Z}Z{\rho}Z,
\EEq
where $p_\Sigma \equiv p_{X}+p_{Y}+p_{Z}$. A special case of (\ref{eq:depolarizationChannel}) is the symmetric depolarization channel, where $p_{X}=p_{Y}=p_{Z}$. The ADC is not sufficient to exactly capture the combined effects of amplitude and phase damping, as no choice of  $p_{X}$, $p_{Y}$, and $p_{Z}$ lead to $\mathcal{E}_{\rm ADC}\left(\rho\right) = \mathcal{E}_{\rm D}\left(\rho\right)$. However, the advantage of the ADC (and more generally the Clifford channel) is that it can be efficiently simulated with a classical computer. Therefore we construct an ADC that approximates (\ref{qubit with decoherence}).

\subsection{Pauli twirl approximation}

We approximate the combined amplitude damping and dephasing with an ADC via {\it twirling} \cite{PhysRevA.72.052326,Emerson28092007,PhysRevA.78.012347,PhysRevA.80.012304,Sarvepalli08052009}. Twirling is used in quantum information to study the average effect of arbitrarily general noise models via their mapping to more symmetric ones. Alternative approximate approaches have also been recently proposed  \cite{2012arXiv1206.5407M,2012arXiv1207.0046G}. 

Using the Kraus matrices (\ref{eq:xdef}), we can rewrite (\ref{qubit with decoherence}) in terms of Pauli matrices as \cite{Sarvepalli08052009},
\BW
\BEq
\label{eq:pauliMatrixExpansion}
\mathcal{E}_{\rm D}\left(\rho\right)=\frac{2-{\gamma}+2\sqrt{1-{\gamma}-\lambda}}{4}~\mathbb{I}{\rho}\mathbb{I}+\frac{\gamma}{4}X{\rho}X+\frac{\gamma}{4}Y{\rho}Y+\frac{2-{\gamma}-2\sqrt{1-{\gamma}-\lambda}}{4}Z{\rho}Z-\frac{\gamma}{4}\mathbb{I}{\rho}Z-\frac{\gamma}{4}Z{\rho}\mathbb{I}+\frac{\gamma}{4i}X{\rho}Y-\frac{\gamma}{4i}Y{\rho}X.
\EEq
Twirling over the Pauli group removes the off-diagonal terms~\cite{PhysRevA.80.012304} from (\ref{eq:pauliMatrixExpansion}), leading to the ADC (\ref{eq:depolarizationChannel}) with error probabilities \cite{Sarvepalli08052009}
\BEq
\label{eq:pixy}
 p_{X} = p_{Y} = \frac{1-e^{-t/T_{1}}}{4} ~~~~~{\rm and}~~~~~ p_{Z} = \frac{1-e^{-t/T_{2}}}{2} - \frac{1-e^{-t/T_{1}}}{4}.
\EEq
If $T_{2}=T_{1}$, the ADC reduces to the symmetric depolarization channel. 
\EW

We refer to the approximate reduction of any quantum channel to the ADC in this manner as the {\it Pauli twirl approximation} (PTA). The PTA corresponds to expanding the Kraus matrices in terms of Pauli matrices (and the identity), performing the Kraus summation, and keeping only terms that are diagonal in the Pauli basis. Equivalently, only the diagonal elements of the $\chi$ matrix in the Pauli basis are retained. Because of its simplicity and wide applicability, we expect the PTA to be a good starting point for refinements that might (approximately) account for the neglected non-diagonal terms. 

\section{physical and logical Errors}
\label{sec:SurfaceCode}

In this section we discuss the assumptions of our error model and the logical error rate in the surface code. We also describe the error correction cycle and review the concept of a distance-dependent error threshold.

\subsection{Error model}
\label{sec:errorModel}

While superconducting qubits promise scalability, they suffer from various error mechanisms caused by gate errors and decoherence \cite{Geller2007Springer2,motzoi2009prl,PhysRevA.85.042321,ghosh2012CZ}. In order to model quantum noise for various surface code architectures we assume that the errors are Markovian (noise affects each individual gate operation independently) and uncorrelated (noise affects each individual qubit separately).

With these assumptions we now describe the dominant error mechanisms relevant for our purpose. We classify these mechanisms as follows:

\begin{enumerate}

\item {\sl Decoherence}. We consider amplitude damping and dephasing as the dominant sources of decoherence, characterized by the relaxation time $T_{1}$ and dephasing time $T_{2}$ of the qubits. Decoherence is introduced here via the PTA as described above, which  allows us to express the single qubit $X$, $Y$, and $Z$ error probabilities as (\ref{eq:pixy}), where $t$ is the operation time. Similarly, with the assumption of uncorrelated errors, one can quantify the error probabilities for various two qubit Pauli channels as
\begin{equation}
\label{eq:depolar2}
\begin{array}{l}
 p_{IX} = p_{IY} = p_{XI} = p_{YI} =   p_{X}(1-p_{X}-p_{Y}-p_{Z}), \\
p_{XX} = p_{XY} = p_{YX} = p_{YY} =   p_{X}p_{Y} ,\\
p_{XZ} = p_{ZX} = p_{YZ} = p_{ZY} = p_{X}p_{Z} ,\\
p_{IZ} = p_{ZI} = p_{Z}(1-p_{X}-p_{Y}-p_{Z}), \\
p_{ZZ} = p_{Z}p_{Z}.
\end{array} 
\end{equation}
Also notice that our assumptions guarantee that any error ($X$, $Y$, or $Z$) in one of the qubits for a two qubit operation can be retrieved when errors on another qubit are traced out; for example $p_{X}=p_{XI}+p_{XX}+p_{XY}+p_{XZ}$. 

\item {\sl Unitary rotation error}. Incorrect unitary operations give rise to a type of intrinsic error. By {\it intrinsic} we mean an error not resulting from noise or decoherence. For single qubit operations, such errors can always be diagonalized in Pauli $X$, $Y$ or $Z$ basis. An estimate suggests that with the use of DRAG pulse shapes \cite{motzoi2009prl}, these errors are ignorable with respect to the intrinsic two-qubit gate errors . Two-qubit gate errors depend on the architecture, and gate protocol.

\item {\sl Leakage}. Leakage is an intrinsic error that populates a quantum state outside of the computational subspace. As far as the single qubit operations are concerned it is possible to suppress leakage below the level of any considerable effect (in comparison to other dominant errors) using quantum control techniques. More quantitatively, it's possible to show that higher-order DRAG pulse is capable to suppress single qubit leakage error below $10^{-8}$ (theoretically) in 5 ns for superconducting qubits \cite{motzoi2009prl}.

\end{enumerate}

In the present analysis, however, our primary focus is to investigate the effect of decoherence on logical error rates and therefore, we do not consider leakage or unitary errors rigorously. Instead, we compute the average intrinsic error of two-qubit gates for the three architectures and distribute it equally to all possible Pauli channels, while decoherence is treated via the PTA.

\subsection{Logical error rate in the surface code}
\label{sec:logicalError}

\begin{figure}[htb]
\includegraphics[angle=0,width=0.88\linewidth]{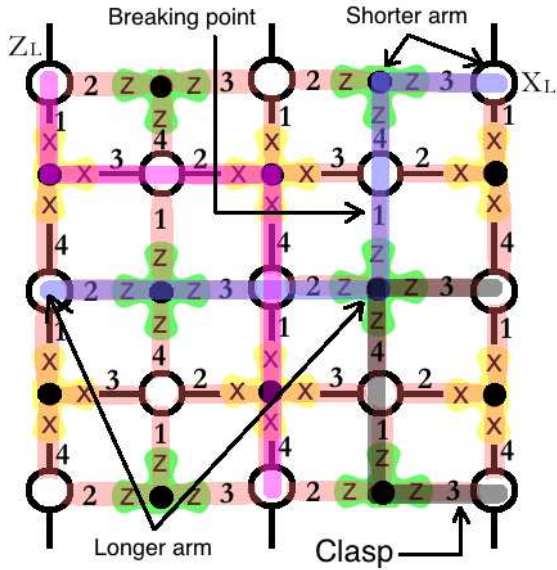}
\caption{(Color online) A schematic diagram of distance-3 surface code is shown. Two possible error chains, $X_{\rm L}$ (purple and horizontal) and $Z_{\rm L}$ (magenta and vertical), are displayed and various terminologies used in this paper are illustrated. Syndrome $Z$ operators are shown in green (labelled by $Z$) and syndrome $X $operators are in yellow (labelled by $X$). An error chain starting and ending at the same boundary is referred to as a `clasp' and is shown in gray color.}
\label{fig:surfaceCode}
\end{figure}

In this section we discuss the use of the surface code as a single-logical-qubit quantum memory and describe the error correction cycle. A distance 3 quantum memory is shown in Fig.~\ref{fig:surfaceCode}. The open circles are data qubits and filled circles are ancillary qubits used for syndrome measurements. A bit-flip on any data qubit results in an eigenvalue change of adjacent $Z$ stabilizers and a phase-flip does the same on neighboring $X$ stabilizers. Therefore, Pauli $X $(bit-flip), $Y $(bit and phase-flip) and $Z$ (phase-flip) errors are detectable (and therefore correctable) by sequential measurements of the stabilizer group generators, unless a misidentification in error-detection leads to the formation of a chain starting from one boundary and ending at another. Such error chains commute with all stabilizers but cannot be written as a product of them and therefore remain undetected. The larger the array (or higher the code distance) the lower the probability of formation of these error chains. 

\begin{figure}[htb]
\includegraphics[angle=0,width=\linewidth]{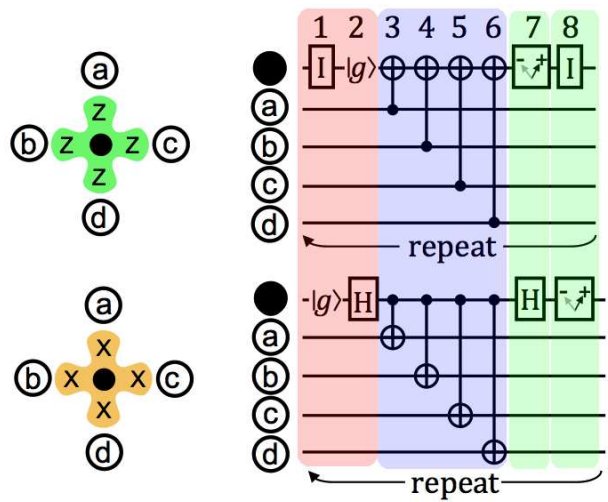}
\caption{(Color online) A schematic diagram of a surface code error correction cycle is shown. The red region (dark gray leftmost region) contains state preparation, the blue region (medium gray middle part) contains four consecutive CNOT operations and the green region (light gray) highlights the measurements of syndrome $Z$ and $X$ qubits.}
\label{fig:circuit}
\end{figure}

Fig.~\ref{fig:circuit} shows the steps that a single surface code error correction cycle is comprised of. The first step is the initial state preparation for the syndrome qubits (state $\ket{0}$ for syndrome $Z$ and $\ket{+}$ for syndrome $X$). While there exists multiple approaches for a qubit state preparation, we here assume that this is done via an ideal projective measurement and a subsequent local rotation ($\sigma^{x}$ or Hadamard), if necessary. The state preparation is followed by four CNOT operations with four adjacent data qubits. The order of these CNOT operations is important and in fact from the reference of a syndrome qubit the clockwise and anti-clockwise orders do not work as they lead to unwanted entanglement among the syndrome qubits \cite{PhysRevA.86.032324}. We here adopt north-west-east-south protocol without any loss of generality. Notice that while for syndrome $Z$ measurements data qubits act as control qubits, for syndrome $X$ measurements data qubits are the targets. These four CNOT operations are followed by measurements for the syndrome $Z$ case and requires a Hadamard operation before syndrome $X$ qubits get measured. Such an error correction cycle can be shown to be equivalent to measuring the four-qubit operators $XXXX$ and $ZZZZ$, and are repeated successively.

The data collected via the measurements of syndrome $Z$ and $X$ qubits at the end of every cycle are stored in a classical computer. A classical minimum-weight perfect matching algorithm is used to match (up to a homology) syndrome events to identify various error chains \cite{dennis:4452,PhysRevA.86.032324}. The most likely logical errors occur when a misidentification by the classical software leads to the formation of an error chain starting from one boundary and ending at another of the same type. Such error chains are referred to as homologically nontrivial error chains and are responsible for logical $X$ or $Z$ operations on the encoded logical qubit. The logical error rate contributed by these error chains can be determined via classical Monte Carlo simulations.

\begin{figure}[htb]
\includegraphics[angle=0,width=\linewidth]{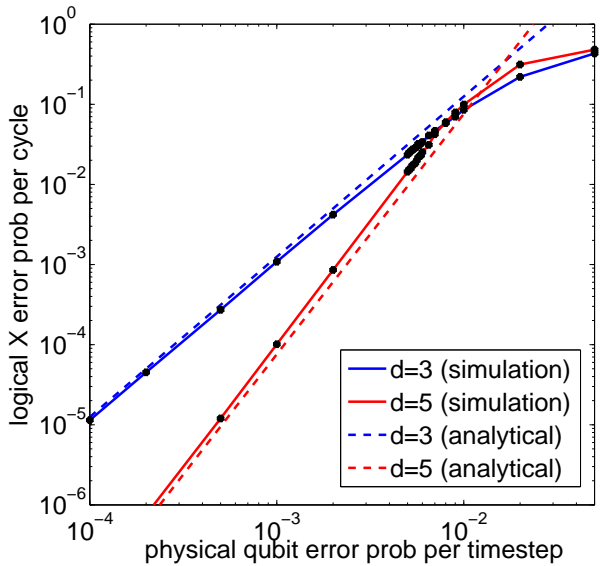}
\caption{(Color online) Plot of analytic estimate of logical $X$ error probability per cycle vs. single physical qubit error probability per timestep. Solid lines denote numerical estimates via Monte-Carlo simulation while dashed lines are obtained from our analytical formula given by Eq.~(\ref{eq:pxl}).}
\label{fig:logicalErrorFormulaComp}
\end{figure}

An analytical leading order estimate of the logical $X$ or $Z$ error rates for an asymmetric depolarization channel error model in a surface code is also derived in Appendix~\ref{app:sec1}, and its performance is compared against the numerical Monte Carlo simulation (as obtained in Ref.~\cite{PhysRevA.86.032324}) in Fig.~\ref{fig:logicalErrorFormulaComp} . As observed in Refs.~\cite{Fowler2011QIC,Fowl12g}, there exists an additional mechanism for logical errors originating from error propagation via CNOT operations---the {\it diagonal} error chains. We neglect such diagonal error chains in the derivation of our analytic formula and therefore it underpredicts the logical error rates. However, a close correspondence between our analytic estimate and numerical simulation is observed for small distance and below threshold, as shown in Fig.~\ref{fig:logicalErrorFormulaComp}, since the contributions from the diagonal error chains are negligible in that regime. Thus the approximate analytic formula is sufficient for the regimes of interest in this work. The convergence of the curves indicate that below the cross-over point, surface code error correction helps as we go from $d=3$ to $d=5$ and above it hurts. We define that transition point as the distance-dependent error threshold.

\section{ARCHITECTURE PERFORMANCE}
\label{sec:architectures}

In this section we perform an analysis of the logical error rate with numerical Monte Carlo simulation (using Autotune \cite{2012arXiv1202.6111F}), and also compare the result to our analytical estimate for the three superconducting architectures. We emphasize that while the numerical Monte-carlo simulation captures all possible error mechanisms, our analytical approach neglects the diagonal error chains as described in Ref.~\cite{Fowler2011QIC,Fowl12g}; it therefore underpredicts the numerical result. However, the analytic formula enables a simple and immediate extension to alternative candidate architectures, error models, and parameter values. Table~\ref{table:model} shows the parameters used to estimate the logical error rate for the three architectures. We assume  tunable transmons for the textbook and Helmer architectures and use two-qubit gate designs that use this tunability. CNOT gates are performed via cross-resonance protocol in the DiVincenzo architecture, which uses transmons operating at the flux sweet spot. Tunable transmons have an additional source of dephasing and therefore we assume $T_{\rm 2}=T_{\rm 1}$ for the textbook and Helmer architectures. State preparation of syndrome qubits is assumed to be done via projective measurement followed by a conditional local rotation (as shown in Fig.~\ref{fig:circuit}) and therefore $t_{\rm QSP}=t_{\rm meas}+t_{\rm loc}$ in Table~\ref{table:model}.

\subsection{Approximate logical error rate}

Here we construct an approximate analytic formula to estimate the  logical error rates below threshold. We use the assumptions of our error model and add the individual error probabilities on data and syndrome qubits for each step to obtain the bit-flip and phase-flip error probabilities {\it per cycle} as
\begin{eqnarray}
\label{eq:textbook}
p_{\rm bf}&=&p_{X}(t_{\rm cycle})+p_{Y}(t_{\rm cycle})+4\frac{8p_{\rm intr}}{15}, \nonumber \\
q_{\rm bf}&=&p_{\rm QSP}+p_{X}(t_{\rm middle})+p_{Y}(t_{\rm middle})+p_{\rm meas}+4\frac{8p_{\rm intr}}{15}, \nonumber \\
p_{\rm pf}&=&p_{Z}(t_{\rm cycle})+p_{Y}(t_{\rm cycle})+4\frac{8p_{\rm intr}}{15}, \nonumber \\
q_{\rm pf}&=&p_{\rm QSP}+p_{Z}(2t_{\rm loc}+t_{\rm middle})+p_{Y}(2t_{\rm loc}+t_{\rm middle}) \nonumber \\
&&+p_{\rm meas}+4\frac{8p_{\rm intr}}{15}, \nonumber \\
\end{eqnarray}
where $t_{\rm middle} \equiv t_{\rm cycle}-(t_{\rm QSP}+t_{\rm loc}+t_{\rm meas})$, $p_{\rm bf}$ and $q_{\rm bf}$ ($p_{\rm pf}$ and $q_{\rm pf}$) are the bit-flip (phase-flip) error rates per cycle in the data qubits and syndrome qubits, respectively. The functions $p(t)$ in (\ref{eq:textbook}) refer to the expressions (\ref{eq:pixy}) evaluated with operation time $t$. Furthermore, $t_{\rm QSP}$ is the time required to complete the initial state preparation for syndrome qubits, and $p_{\rm QSP}$ is the error probability that a wrong state is prepared. $p_{\rm intr}$ is the intrinsic error of a CNOT gate averaged over the Hilbert space of all input states, and $p_{\rm meas}$ is the error probability that a wrong eigenvalue is reported in the readout process. Note that the intrinsic gate error (described by $p_{\rm intr}$) is assumed to be equally distributed over all 15 two-qubit Pauli errors and therefore the probability of a bit flip (occurs with $X$ or $Y$ errors) or phase flip (occurs with $Z$ or $Y$ errors) of any qubit during CNOT due to the intrinsic error is $8p_{\rm intr}/15$. When we add each probability we are ignoring all higher-order contributions and also the coherence of these error mechanisms (although our numerical simulation takes the higher-order effects into account). We use these bit and phase flip probabilities in Eq.~(\ref{eq:pxl}) and Eq.~(\ref{eq:pzl}) of Appendix \ref{app:sec1} to obtain analytical estimates of logical $X$ and $Z$ error rates.

\BW
\begin{center}
\begin{table}[htb]
\caption{Parameters assumed for the three fault-tolerant architectures.}
  \begin{tabular}{| c | c | c | c | c |}
  \cline{1-5}
& & \multicolumn{3}{ c |}{architectures} \\
    \cline{3-5}
    quantity & description & textbook & Helmer & DiVincenzo \\ \hline
    $T_{1}$ & qubit relaxation time & 1-10 ${\mu}s$ & 1-10 ${\mu}s$ & 1-40 ${\mu}s$\\ \hline
    $T_{2}$ & qubit dephasing time & $T_{1}$ & $T_{1}$ & $2T_{1}$\\ \hline
    $t_{\rm QSP}$ & state preparation time & 40 ns & 40 ns & 40 ns \\ \hline
    $t_{\rm loc}$ & local rotation time & 5 ns & 5 ns & 5 ns \\ \hline
    $t_{\rm meas}$ & measurement time & 35 ns & 35 ns & 35 ns \\ \hline
    $t_{\rm CNOT}$ & CNOT gate time & 21 ns & 20 ns & 20 ns \\ \hline
    $t_{\rm cycle}$ & time duration of a single cycle & 164 ns & 160 ns & 400 ns \\ \hline
    $p_{\rm intr}$ & leakage probability for CNOT & $10^{-4}$ & $10^{-3}$ & $10^{-3}$ \\ \hline
    $p_{\rm meas}$ & measurement error probability & $10^{-2}$ & $10^{-2}$ & $10^{-2}$ \\ \hline
    $p_{\rm QSP}$ & state preparation error probability & $10^{-2}$ & $10^{-2}$ & $10^{-2}$ \\ \hline
  \end{tabular}
\label{table:model}
\end{table}
\end{center}
\EW

\subsection{Textbook architecture}

The textbook architecture consists of a two-dimensional square lattice (as shown in Fig.~\ref{fig:textbook}) of superconducting qubits---tunable transmons---with nearest-neighbor tunable couplings having infinite on-off ratio. The CNOT operations in this architecture are performed using the protocol discussed in Ref.~\cite{ghosh2012CZ}. We assume that the idle data qubit frequencies are 6 GHz and syndrome qubit frequencies are 8 GHz. The optimal parameters for a CNOT operation, shown in Table \ref{table:twoQubitModel}, are determined in Appendix \ref{app:sec2} by modeling amplitude and phase damping. As mentioned earlier, $T_{\rm 1}=T_{\rm 2}$ is assumed for tunable transmons, as they have an additional source of dephasing that degrades their $T_{\rm 2}$. 

\begin{figure}[htb]
\centering
\subfloat[]{\includegraphics[angle=0,width=\linewidth]{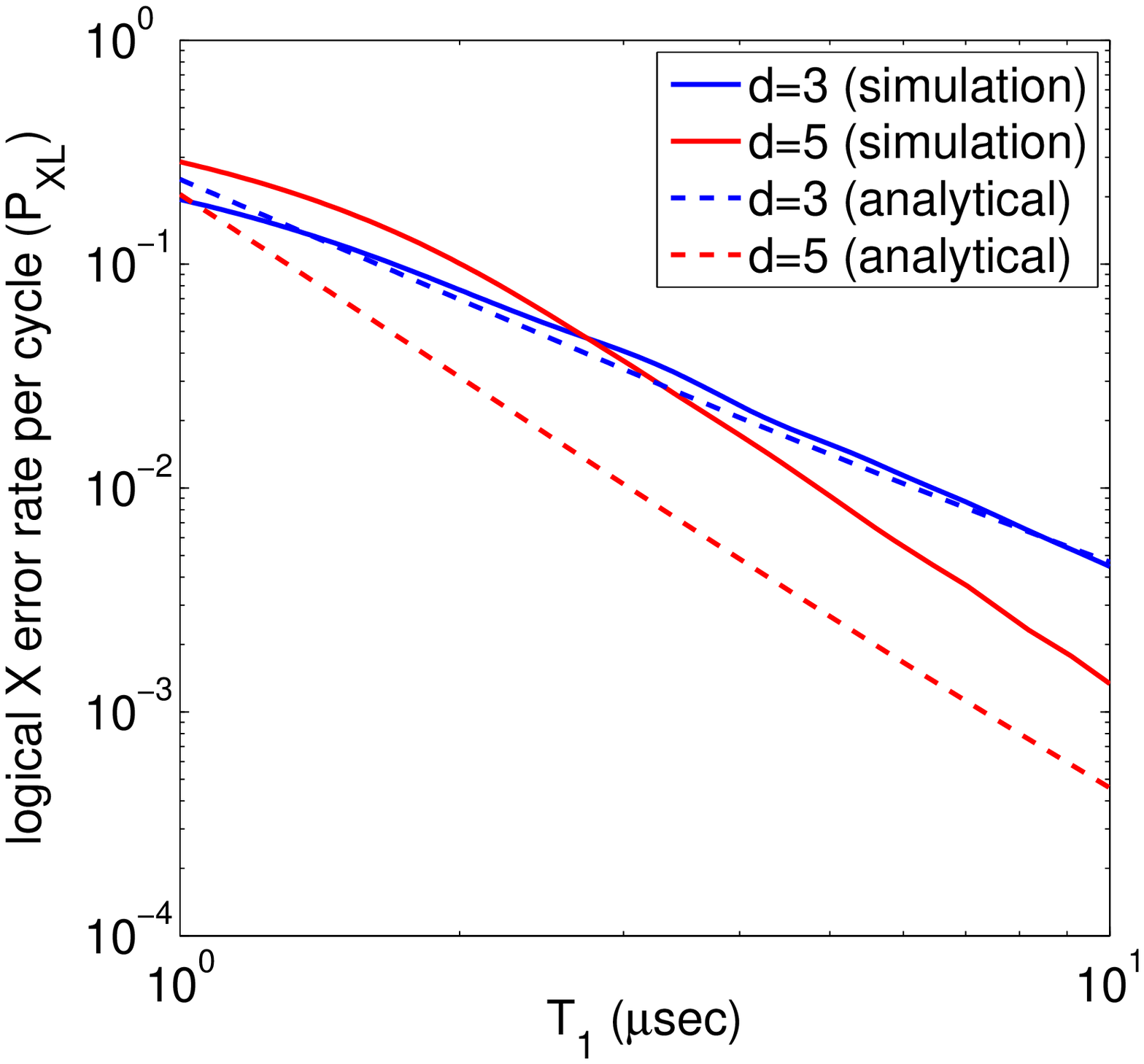}} \\
\subfloat[]{\includegraphics[angle=0,width=\linewidth]{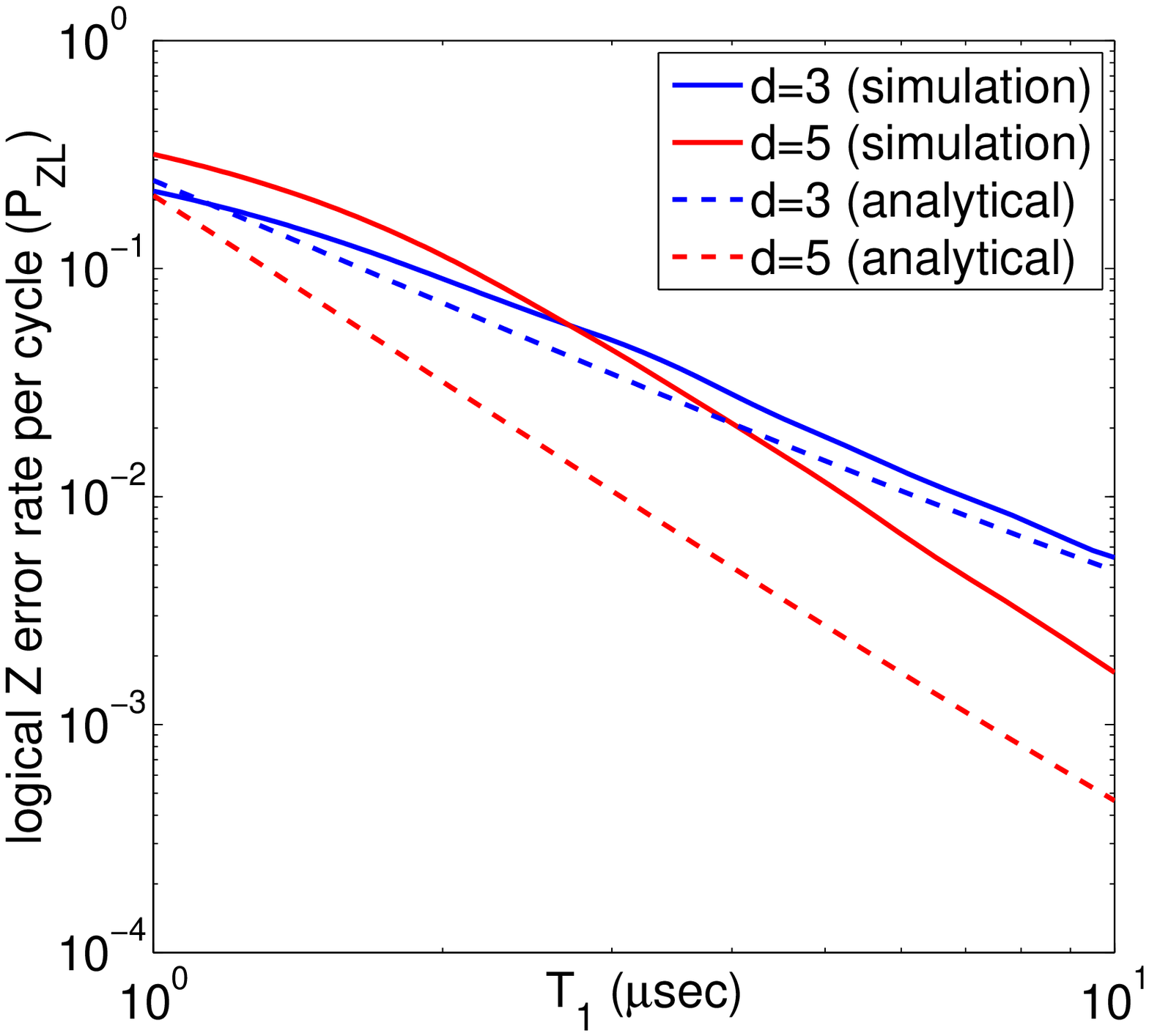}}
\caption{(Color online) Logical $X$ and $Z$ error rate per cycle is shown as a function of coherence time $T_1$ for the textbook architecture. Plots for $d=3$ are shown in blue and those for $d=5$ are shown in red.}
\label{fig:textbook_error}
\end{figure}

We use Eq.~(\ref{eq:textbook}) along with Eq.~(\ref{eq:pxl}) and Eq.~(\ref{eq:pzl}) to compute the logical $X$ and $Z$ error rates ($P_{\rm XL}$ and $P_{\rm ZL}$) for the textbook architecture. For the numerical Monte Carlo simulation, we use Autotune \cite{2012arXiv1202.6111F} to simulate the circuit shown in Fig.~\ref{fig:circuit} for every syndrome qubit.

In Fig.~\ref{fig:textbook_error} we show the graphs (both analytical and numerical) of the logical $X$ and $Z$ error probabilities per cycle with $d=3$ and $d=5$ codes, versus the relaxation time $T_{\rm 1}$. Note that for $d=3$ our analytic formula closely reproduces the numerical simulation, while for $d=5$ it underpredicts as we expect. From the numerical plots we observe that the threshold is at $\approx 2.6$ ${\mu}s$, where all other parameters are kept fixed as listed in Table \ref{table:model}. This result signifies that if we construct this architecture with qubits having $T_{\rm 1}$ (or $T_{\rm 2}$) more than $\approx 2.6$ ${\mu}s$, then surface code error correction helps as we increase the distance from $d=3$ to $d=5$; otherwise it hurts.

\subsection{Helmer architecture} 
\label{sec:HelmerArchitecture}

In this section we discuss the architecture proposed by Helmer \textit{et al.} \cite{0295-5075-85-5-50007}, where superconducting qubits are arranged in a two-dimensional square lattice and each qubit is coupled to one horizontal and one vertical cavity as shown in Fig.~\ref{fig:Helmer}. The rectangular blocks (horizontal and vertical) are cavities, circles represent qubits and the colors denote their idle (between gate) frequencies. As pointed out in Ref.~\cite{0295-5075-85-5-50007}, the minimum frequency range required to allocate the frequencies of all qubits in this architecture is proportional to square root of the number of qubits. While this architecture is not scalable, it is suitable for implementing the distance 3 and 5 surface code, which is a main focus here.

The CNOT gates are performed between a pair of adjacent qubits by tuning them into mutual resonance and waiting for a while somewhere near cavity frequency and thereby utilizing the effective flip-flop interaction between qubits. The waiting time for this gate is inversely proportional to the magnitude of the effective flip-flop interaction strength and for parameters used in Ref.~\cite{0295-5075-85-5-50007} we estimate $t_{\rm CNOT} \approx 20$ ns for this protocol. The dominant source of intrinsic errors for such a CNOT emerges from the higher-order Landau-Zener transitions during tuning and detuning and are estimated to be in the order of $10^{-3}$ \cite{0295-5075-85-5-50007}. As specified earlier, the parallel CNOT operations involving the same resonator also cost fidelity due to the higher-order couplings in this architecture. However, in the low distance limit we assume that the total intrinsic error is bound by the fixed (distance-independent) value mentioned above. These parameters are shown in Table \ref{table:model} and used to estimate the logical error probability per cycle for this architecture.

\begin{figure}[htb]
\centering
\subfloat[]{\includegraphics[angle=0,width=\linewidth]{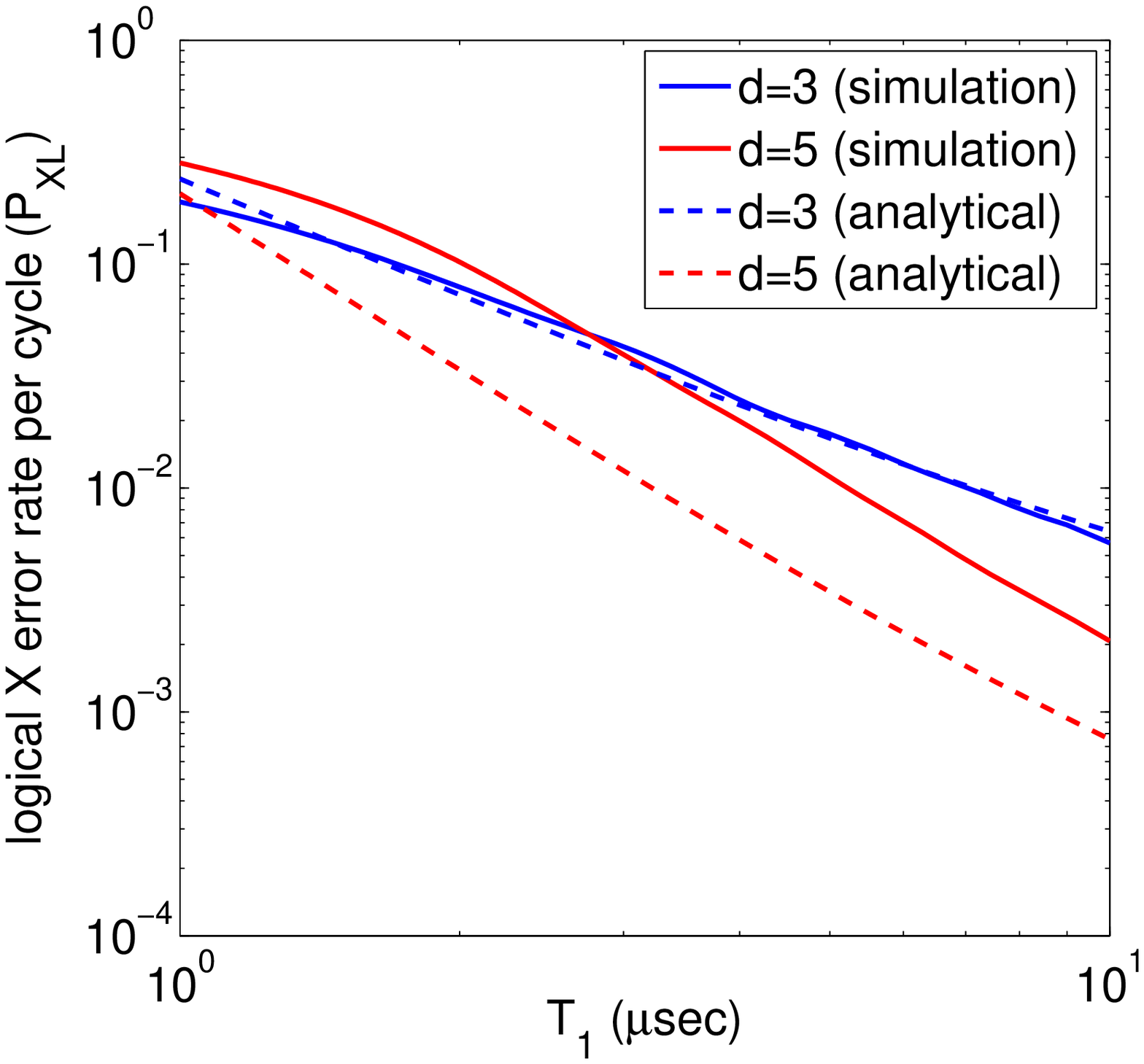}} \\
\subfloat[]{\includegraphics[angle=0,width=\linewidth]{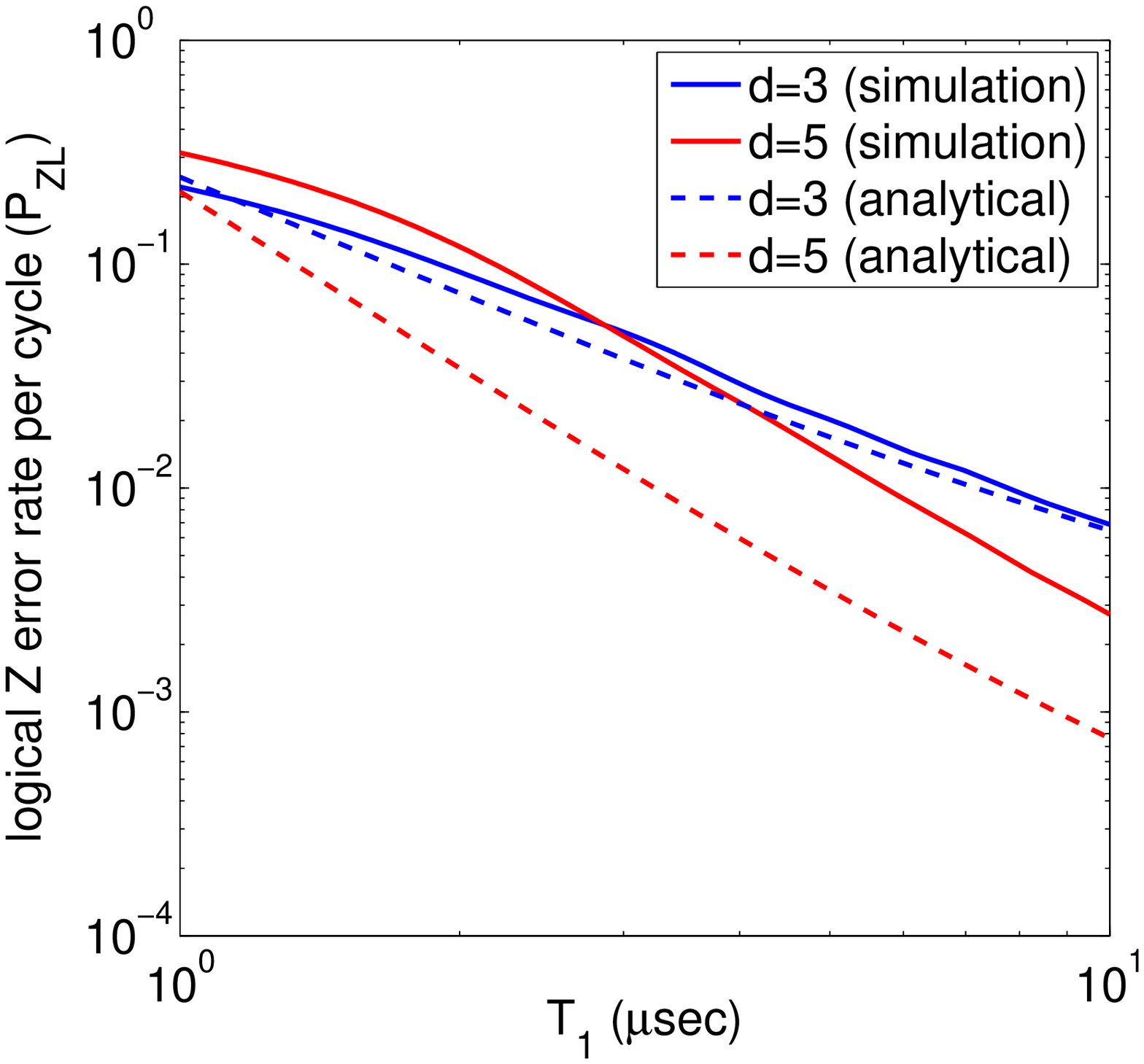}}
\caption{(Color online) Logical $X$ and $Z$ error rate per cycle is shown as a function of coherence time $T_1$ for the Helmer architecture. Plots for $d=3$ are shown in blue and those for $d=5$ are shown in red.}
\label{fig:helmer_error}
\end{figure}

The bit-flip and phase-flip error probabilities per cycle for data and syndrome qubits in this architecture are given by  (\ref{eq:textbook}) with $t_{\rm cycle}=160$ ns and $t_{\rm middle}=(4 \times 20)$ ns $=80$ ns. With a similar analysis we obtain Fig.~\ref{fig:helmer_error}, which shows the plots of logical $X $and $Z$ error probabilities per cycle for $d=3$ and 5 error correction with respect to $T_{\rm 1}$, and we observe that the threshold is at $\approx 2.8$ ${\mu}s$.

\subsection{DiVincenzo architecture}
\label{sec:IBMarchitecture}

Here we analyze the architecture (shown in Fig.~\ref{fig:IBM}) proposed by DiVincenzo \cite{1402-4896-2009-T137-014020}, in which each qubit is dispersively coupled to two resonators, while each resonator couples four such qubits. In this architecture every data or syndrome qubit consists of four physical qubits where one of them is primary and the remaining three act as ancillary qubits. The CNOT operations in this architecture are performed via the virtual cross resonance protocol where qubits always remain dispersively coupled to the resonators while microwaves drive the population transition between two qubits \cite{PhysRevLett.107.080502,2012arXiv1202.5344C,PhysRevB.81.134507}. Notice that this architecture is fully scalable and the frequency allocation does not depend on the number of qubits. 

\begin{table} 
\caption{Time duration for each step in the error-correction cycle for DiVincenzo architecture.}
\begin{tabular}{|c|c|}
\hline operation & time duration \\ 
\hline state preparation & 40 ns \\ 
\hline first CNOT (north) & 100 ns \\ 
\hline second CNOT (west) & 60 ns \\ 
\hline third CNOT (east) & 60 ns \\ 
\hline fourth CNOT (south) & 100 ns \\ 
\hline local rotation plus readout & 40 ns \\ 
\hline 
\end{tabular} 
\label{table:IBM}
\end{table}

We first estimate the time required to complete a single surface code cycle in this architecture. As mentioned earlier, for each block one out of four qubits acts as a principal qubit and without loss of generality we assume the eastern qubit to be the principal one for every block. Table~\ref{table:IBM} shows the time required for each individual step in this architecture. The state preparation and read out takes 40 and 35 ns, respectively, as for previous architectures. The first CNOT is performed between a syndrome block and its north data qubit block and this is performed by doing a CNOT between the eastern qubit of the syndrome block and the western qubit of the data block. This CNOT must be accompanied by pre- and post-SWAP operations in the data block where the quantum state of the eastern qubit is transferred to the western one. As discussed, the CNOT operations are performed via the cross-resonance protocol and we assume the gate time for such a CNOT to be $\approx 20$ ns \cite{PhysRevB.81.134507}. SWAP operations between two qubits coupled via resonator is also assumed to be performed in 20 ns. The intrinsic error $p_{\rm intr}$ for such CNOT gates is estimated to be in the order of $10^{-3}$ \cite{2012arXiv1202.5344C}. These results give us the time durations required for each step in the error correction cycle, shown in Table~\ref{table:IBM}. These estimate the duration of a single cycle in this architecture to be 400 ns long.

\begin{figure}[htb]
\centering
\subfloat[]{\includegraphics[angle=0,width=\linewidth]{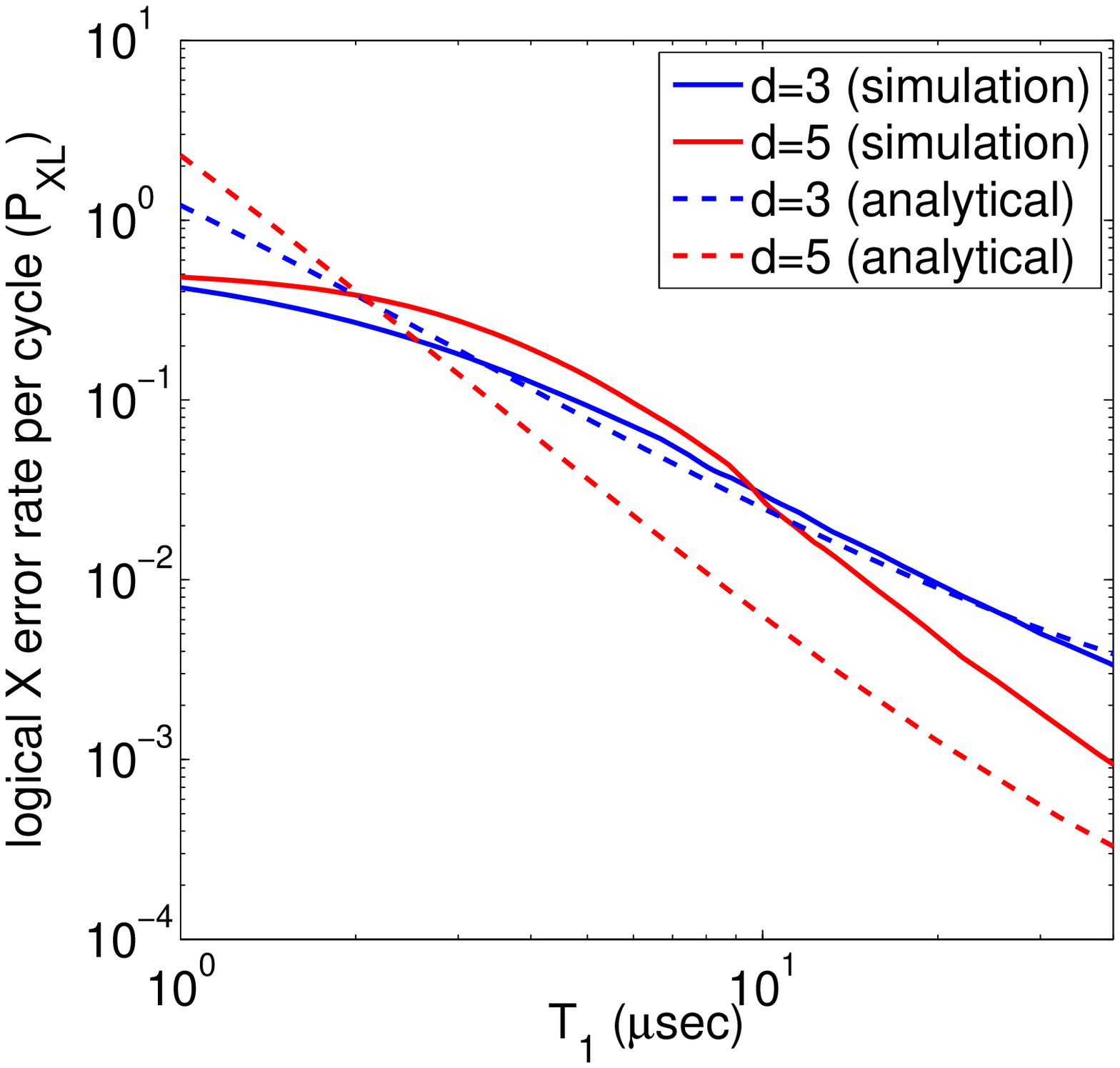}} \\
\subfloat[]{\includegraphics[angle=0,width=\linewidth]{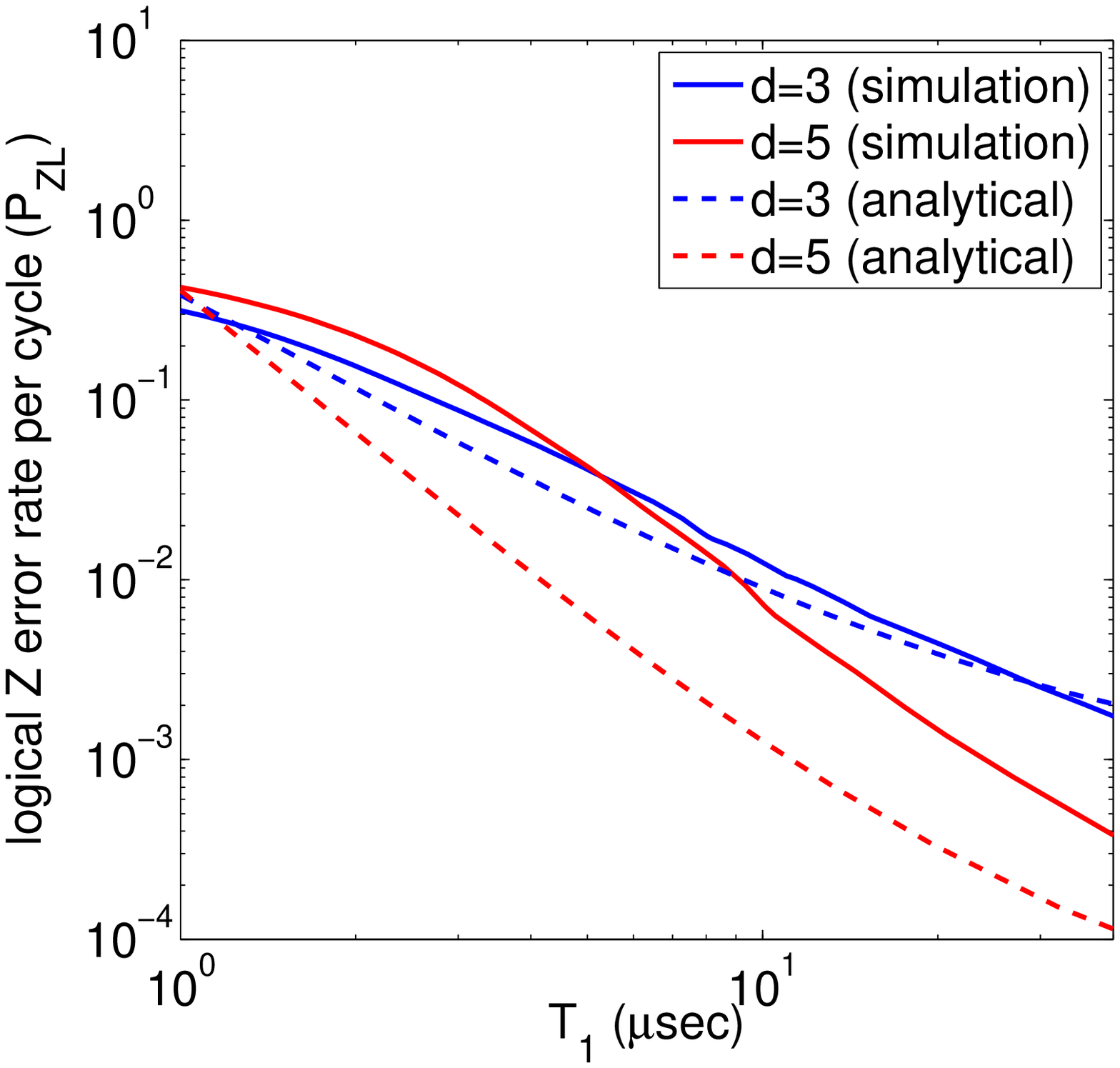}}
\caption{(Color online) Logical $X$ and $Z$ error rate per cycle is shown as a function of coherence time $T_1$ for the DiVincenzo architecture. Plots for $d=3$ are shown in blue and those for $d=5$ are shown in red.}
\label{fig:divincenzo_error}
\end{figure}

Following the same argument as in the textbook architecture and using (\ref{eq:textbook}) for the bit-flip and phase-flip error probabilities with $t_{\rm cycle}=400$ ns and $t_{\rm middle}=(100+60+60+100)$ ns $=320$ ns, we compute logical $X$ and $Z$ error rates. Fig.~\ref{fig:divincenzo_error} shows the total logical $X$ and $Z$ error probabilities per cycle for $d=3$ and $5$. Note that the condition, $T_2=2T_1$, leads to $p_X+p_Y \approx 2(p_Z+p_Y)$ (assuming $T_{\rm 1,2} \gg t_{\rm cycle}$), which means that the bit-flip error rate is almost twice larger than the phase-flip error rate. Since, logical $X$ error rate mostly depends on bit-flip probability and logical $Z$ on phase-flip, we expect $P_X > P_Z$ for this case. This asymmetry between logical $X$ and $Z$ error rates imply a larger error threshold for logical $X$ error in comparison to logical $Z$. We observe from our numerical simulation that logical $Z$ errors can be suppressed if $T_1 > 5 ~{\mu}s$, while in order to suppress logical $X$ errors we need $T_1 > 10 ~{\mu}s$, which is consistent with the above argument.

\subsection{Other possible architectures} 

\begin{figure}[htb]
\centering
\subfloat[]{\includegraphics[angle=0,width=0.48\linewidth]{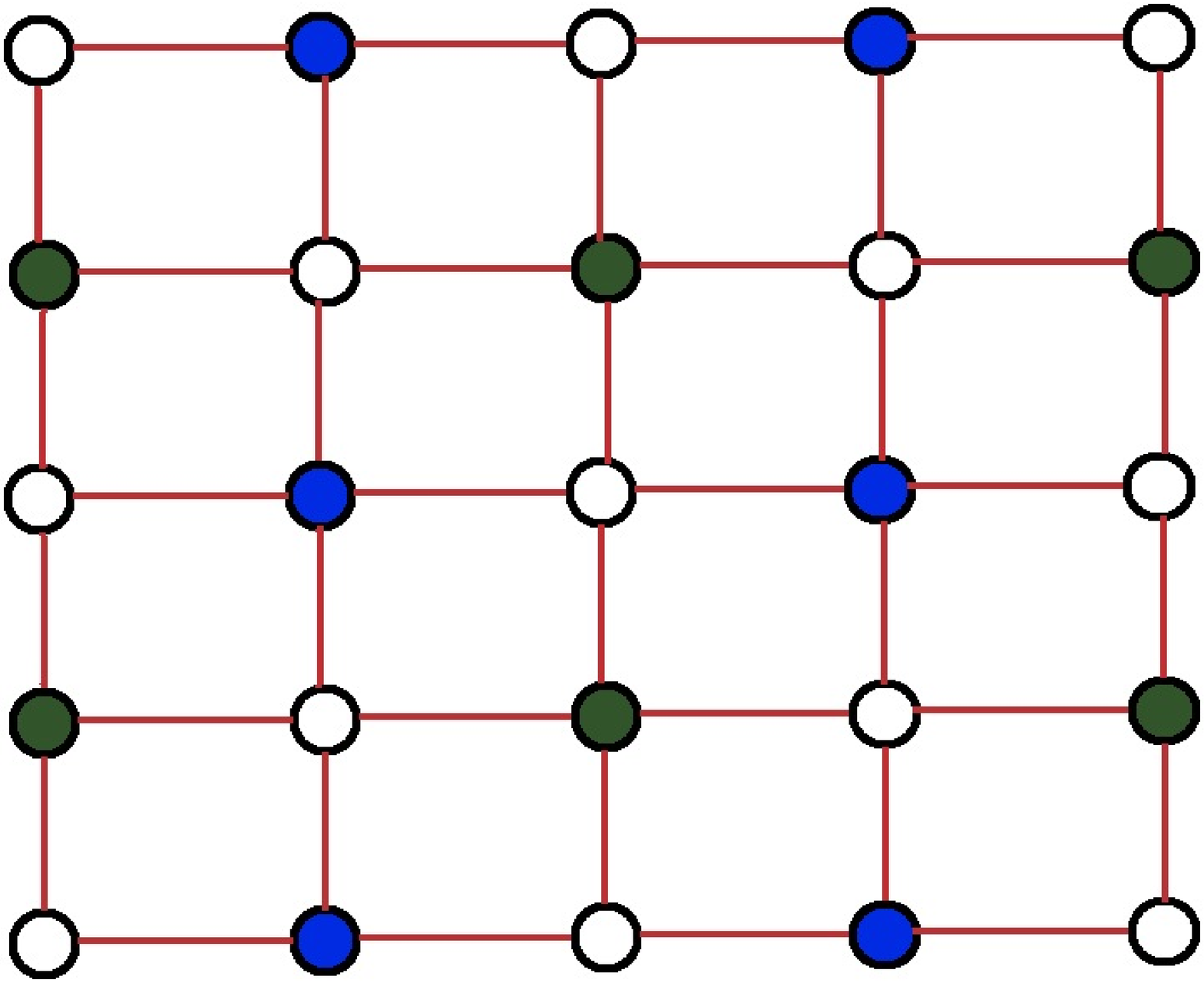}}
\subfloat[]{\includegraphics[angle=0,width=0.515\linewidth]{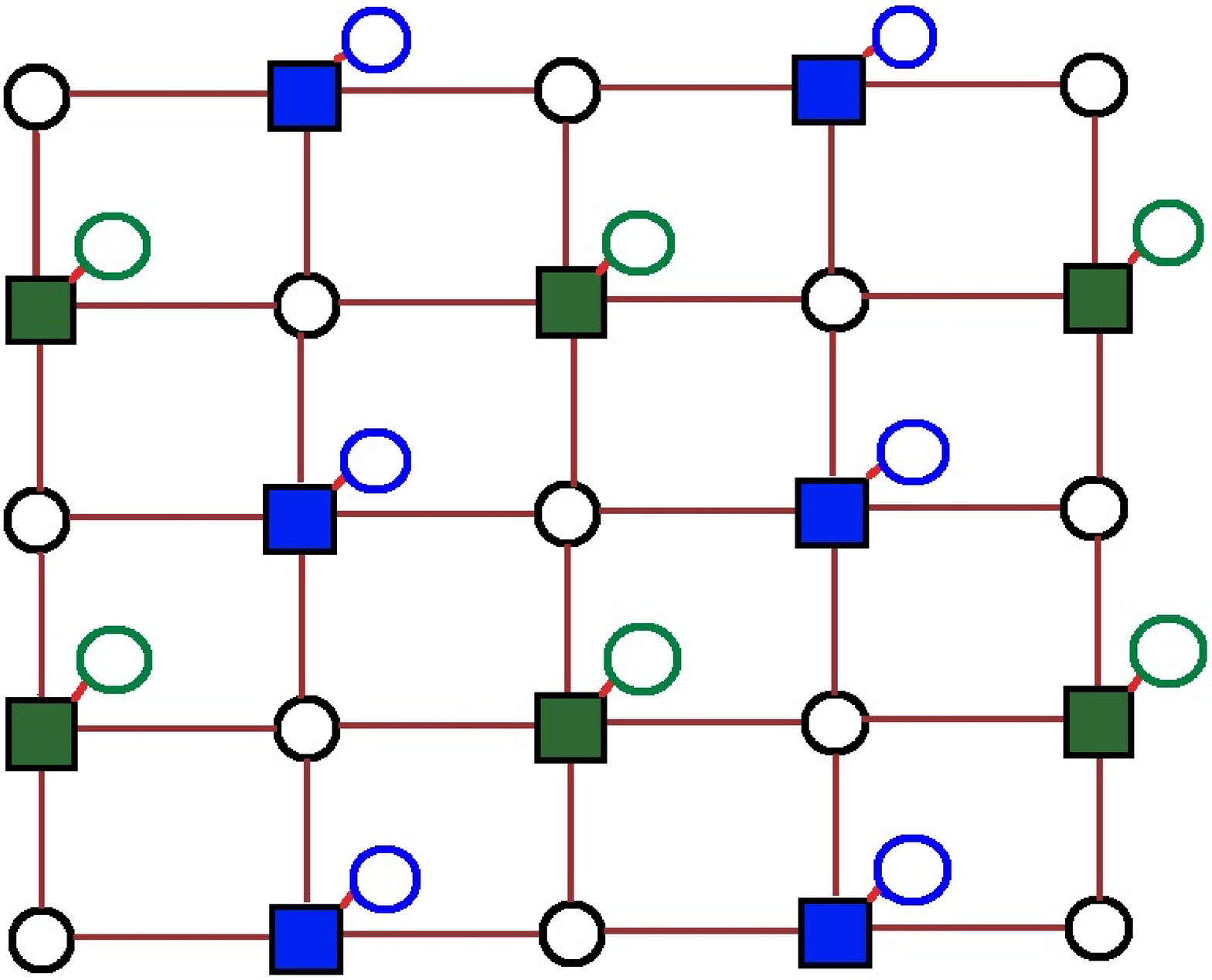}} \\
\subfloat[]{\includegraphics[angle=0,width=0.5\linewidth]{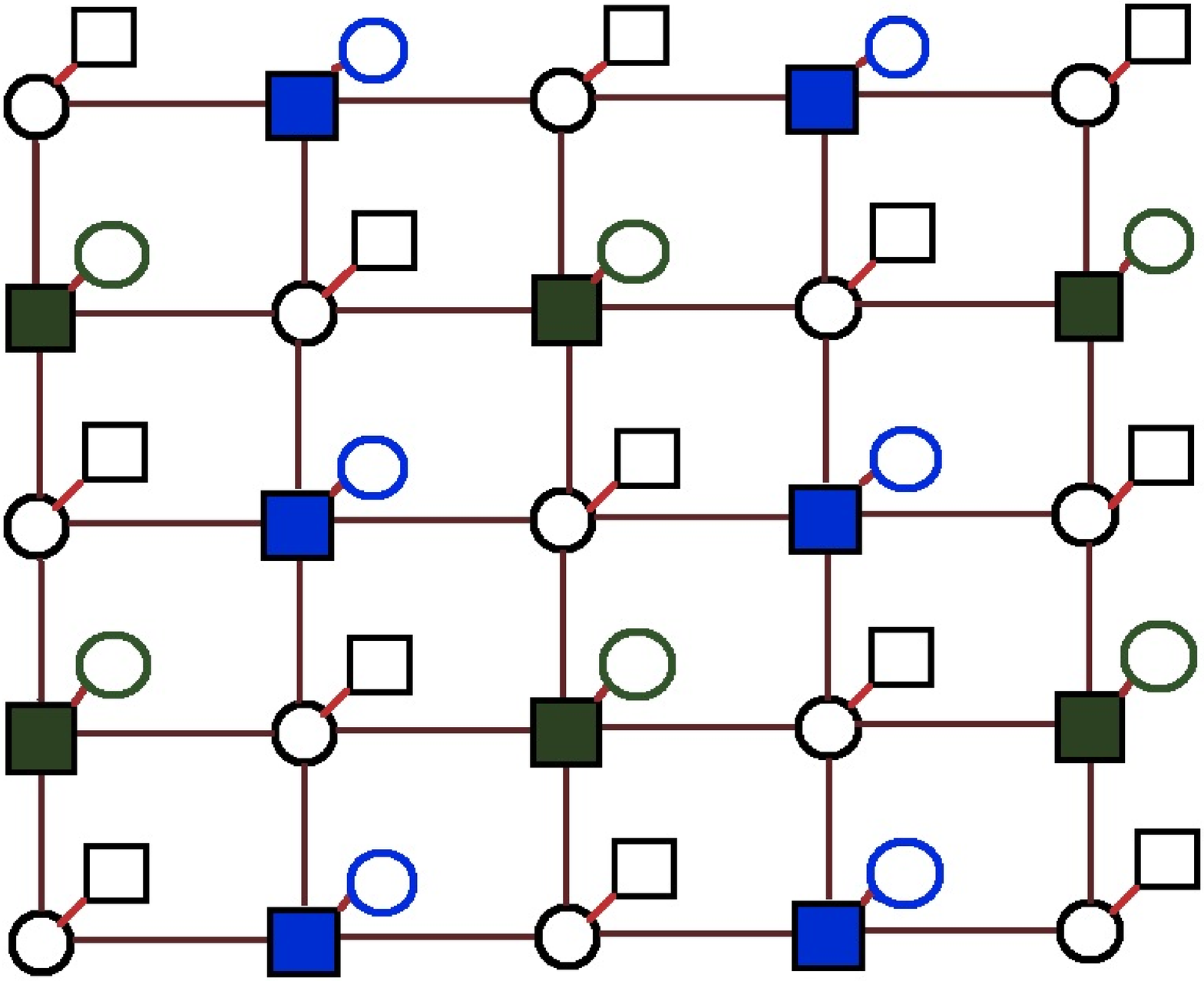}}
\subfloat[]{\includegraphics[angle=0,width=0.5\linewidth]{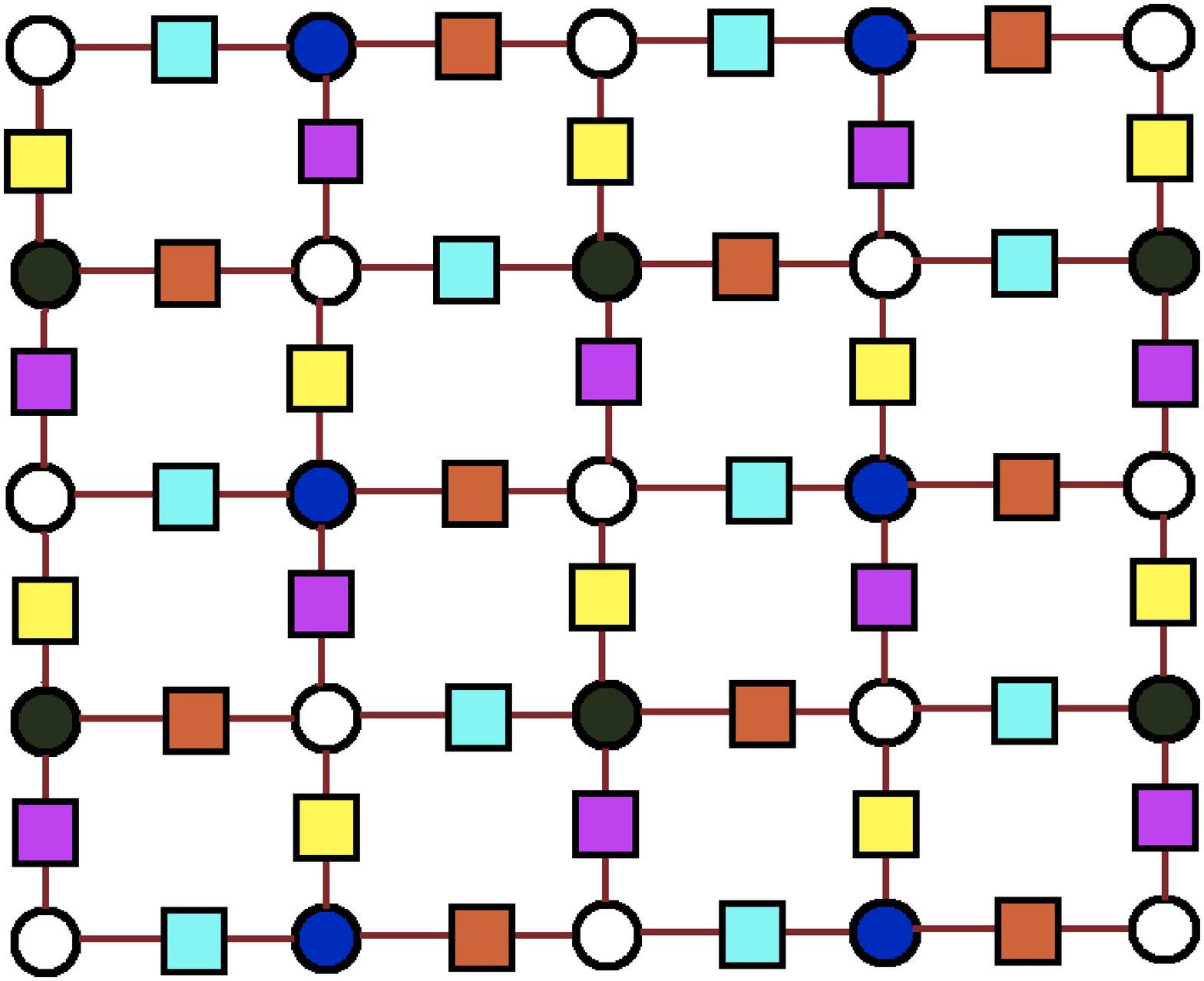}} 
\caption{(Color online) Various possible fixed coupling-based architectures are shown for d=3 surface code. The circles denote qubits, squares denote resonators and various colors (grayscales) denote a possible frequency allocation. (a) An architecture where superconducting qubits are arranged in a two-dimensional square lattice each coupled to its nearest neighbor with fixed couplers. (b) An architecture where superconducting qubits are used for data qubits and resonators for syndrome qubits coupled via fixed couplers. Each resonator is also coupled to another qubit required for read out. (c) Same as architecture (b) except for the fact that each qubit is also coupled to another resonator used as its memory. (d) In this architecture each qubit in a two-dimensional square lattice is coupled to its nearest neighbor via a resonator.}
\label{fig:fixedArch}
\end{figure}

We also discuss some other possible architectures based on fixed coupling elements, as shown in Fig.~\ref{fig:fixedArch}. As per our convention, the squares denote resonators, circles denote qubits, and solid lines denote fixed couplings. For gate protocols (CNOT or SWAP) that require tuning and detuning qubits in and out of resonance, the greatest challenge is the frequency allocation such that first-order Landau-Zener transitions can be avoided. We observe that with DC control-based gate protocols \cite{StrauchPRL03,ghosh2012CZ}, none of these architectures can avoid first order Landau-Zener transitions. This fact is an inherent property of the topology of these architectures. However, we note that with microwave-control-based gate protocols (for example, cross-resonance), these unwanted transitions can be avoided.

The crucial role of Landau-Zener transition on the error mechanisms for these architectures motivates us to estimate this error: For any two level system Landau-Zener formula predicts the diabatic transition probability as
\BEq
\label{eq:LZ}
P_{\rm LZ}=\exp\left({-2{\pi}\frac{g^{2}}{{\hbar}|\dot{\epsilon}(t^{*})|}}\right),
\EEq
where $g$ is the coupling between the levels, $\epsilon(t)$ is the time-dependent energy level separation and $t^{*}$ is the time when the levels are in resonance. Assuming parameters relevant for superconducting architectures ($g \approx 45\, {\rm MHz}$ and $\dot{\epsilon}(t^{*}) \approx 2\, {\rm GHz/ns}$), we obtain a Landau-Zener transition error $1-P_{\rm LZ}$ of about 4\%. This error is unacceptably large. We do not attempt here to perform a more quantitative analysis of logical error rate for these architectures. However, we emphasize that while microwave control-based gate protocols may prove to be useful for these cases, such gate operations have not yet been analyzed in the context of these designs.

\section{Conclusion}
\label{sec:conclusion}

\begin{center}
\begin{table}[htb]
\caption{ Fault-tolerant $T_{\rm 1}$ thresholds for the three architectures studied in this work.}
  \begin{tabular}{| c | c | c |}
  \cline{1-3}
& \multicolumn{2}{ c |}{ $T_{\rm 1}$ threshold} \\
    \cline{2-3}
    architecture & logical $X$ error & logical $Z$ error \\ \hline 
    textbook & 2.6 ${\mu}s$ & 2.6 ${\mu}s$ \\ \hline
    Helmer & 2.8 ${\mu}s$ & 2.8 ${\mu}s$ \\ \hline
    DiVincenzo & 10 ${\mu}s$ & 5 ${\mu}s$ \\ \hline
  \end{tabular}
\label{table:result}
\end{table}
\end{center}

We have investigated the logical error rate and fault-tolerant error threshold for three superconducting surface code implementations. While the coherence time has been improving over the past few years for superconducting qubits, we discuss here the minimum coherence time required to achieve error correction. The logical error rate for $d=3$ and $5$ is computed as a function of qubit coherence time and the threshold is found to be dependent on the architecture, error model, and assumed gate protocol. Table~\ref{table:result} shows our main results. These error thresholds are within reach of current state-of-the-art superconducting circuit designs. The operation time requirements for qubit state preparation and readout are, however, yet to be achieved experimentally to the accuracy assumed in this work. Our analysis can be extended to the future surface code architectures. As mentioned earlier, the effect of decoherence on the logical error rate is a primary focus in this work, and our error models neglect various higher-order and unintended stray couplings between qubits. Exploring the effect of these factors is a possible direction of future research.

\begin{acknowledgments}
This work was supported by IARPA under ARO Grant No. W911NF-10-1-0334. It is a pleasure to thank John Martinis for many useful discussions and suggestions.
\end{acknowledgments}

\appendix
\section{Derivation of approximate surface code logical error rate}
\label{app:sec1}

In this section we derive the logical error per qubit per cycle as a function of the single qubit error rates, to leading order. Our derivation here does not include the ``diagonal" error propagation via CNOT gates \cite{Fowler2011QIC,Fowl12g} and therefore underestimates the logical error rates. Logical error rates for $X$ and $Z$ errors per cycle ($P_{\rm XL}$ and $P_{\rm ZL}$ respectively) are defined as the probability of formation of an $X$ or $Z$ error chain in the surface at the end of a single cycle. We consider the logical $X$ error first, and the expression for the logical $Z$ error follows from a similar combinatorial argument. Suppose $p_{\rm bf}$ and $q_{\rm bf}$ are bit-flip error probabilities (per cycle) in the data and syndrome qubits, respectively. The dominant error mechanism emerges from the fact that $(d+1)/2$ errors either get misidentified as $(d-1)/2$ errors (with 100\% probability) or as a different arrangement of $(d+1)/2$ errors (with 50\% probability), thereby producing an error chain after attempting error correction. Such a process can happen in three ways.

\textit{Case 1.}  The most natural error chain happens when there are $(d+1)/2$ data-qubit bit-flip errors in a single row of a distance-$d$ surface. These $(d+1)/2$ error locations can be chosen out of $d$ locations in ${d \choose \frac{d+1}{2}}$ ways, and such an error chain may occur in any one of the $d$ rows, leading to
 \BEq
 P_{\rm XL}^{(1)}=d \, {d \choose \frac{d+1}{2}} \, p_{\rm bf}^{\frac{d+1}{2}}.
 \EEq
Note that the chance of misidentification of these $(d+1)/2$ errors is 100\% for this case because the classical error detection software is based on  minimal-weight perfect matching. This expression was previously derived in Ref.~\cite{PhysRevA.86.032324}.

\textit{Case 2.} In this case $(d+1)/2$ errors occur in two consecutive rows, as shown in Fig.~\ref{fig:surfaceCode}. We refer to such an error chain as a `broken' error chain and call the point where the chain changes its row as the `breaking point' (shown in Fig.~\ref{fig:surfaceCode}). In order to estimate this case correctly one needs additional care with error-chains starting from one boundary and ending at the same boundary in a different row. We refer to such an error chain as a `clasp' (shown in Fig.~\ref{fig:surfaceCode}). Notice that clasps are homologically trivial and therefore should not be considered as a source of logical error. In order not to count these clasps, we classify this case into two mutually exclusive and exhaustive (to leading order) subcases: ${\rm i\rangle}$ when errors occur in horizontal links of a surface code lattice and ${\rm ii\rangle}$ when there are no errors on horizontal links. Also, observe that chains with errors in more than one horizontal links contribute to a higher-order process and are therefore excluded from our leading order analysis. If we think that the horizontal link with error divides a row into shorter and longer arms (also shown in Fig.~\ref{fig:surfaceCode}), then for subcase i, the number of ways an error chain is formed (${\cal W^{\rm 1}}$) is constrained by the condition that all sites of the shorter side cannot be filled with errors for any error chain since in that subcase one would be constructing a clasp. Satisfying this condition, for a given orientation and a specific pair of adjacent rows, we obtain,
\begin{eqnarray}
{\cal W^{\rm 1}} &=& \underbrace{(d-1){d \choose \frac{d-1}{2}}}_\text{all possible chains}-\underbrace{2\sum_{r=1}^{\frac{d-1}{2}}{d-r \choose \frac{d-1}{2}-r}}_\text{clasps} \nonumber \\
&=& \frac{d^{2}-1}{d+3}{d \choose \frac{d-1}{2}}. 
\end{eqnarray}
For subcase ii all the single physical qubit errors are distributed among vertical links in two adjacent rows. In this subcase, for a given distribution of single qubit errors, in order not to overcount the homotopic error chains one needs to adopt a convention to place the breaking point. Without loss of any generality, we adopt the convention that the breaking point for this subcase is always placed right next to the rightmost error on the lower arm. Such a convention prevents overcounting of homotopic error chains. The remaining condition one needs to satisfy for this subcase is not to place all single qubit errors on the longer arm of the error chain. This condition prevents us from overcounting case 1. Satisfying these conditions, we find the number of ways an error chain is formed (${\cal W^{\rm 2}}$) for subcase ii as,
\begin{eqnarray}
{\cal W^{\rm 2}} &=& \underbrace{(d-1){d-1 \choose \frac{d-1}{2}}}_\text{all possible chains}-\underbrace{\sum_{r=2}^{\frac{d+1}{2}}{d-r \choose \frac{d-1}{2}}}_\text{\textit{Case-1} chains} \nonumber \\
&=& \frac{d-1}{2}{d \choose \frac{d-1}{2}}. 
\end{eqnarray}
Combining these results we obtain the logical $X$ error probability per cycle for case 2 as
\begin{eqnarray}
\label{eq:pxl2}
  P_{\rm XL}^{(2)} &=& \frac{1}{2}2(d-1)\left[{\cal W^{\rm 1}}+{\cal W^{\rm 2}}\right]p_{\rm bf}^{\frac{d+1}{2}} \nonumber \\
  &=& (d-1)\frac{(3d+5)(d-1)}{2(d+3)}{d \choose \frac{d-1}{2}}p_{\rm bf}^{\frac{d+1}{2}} .
\end{eqnarray}
In the first line, the factor of 2 comes from the orientation (bottom-left to top-right or top-left to bottom-right) of the error chain, the factor of 1/2 denotes the fact that the classical error detection software misidentifies such an error chain with a 50\% probability, and $d-1$ corresponds to the number of adjacent pair of rows in a distance-$d$ code.

\begin{figure}[htb]
\includegraphics[angle=0,width=\linewidth]{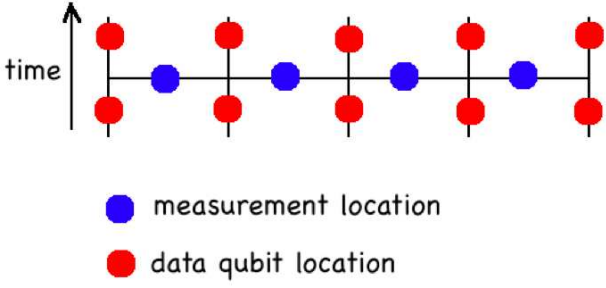}
\caption{(Color online) Data qubits (red filled circles) of a single row in a distance-5 surface code is shown in two subsequent time slices. The blue filled circles denote measurement locations. An error in any measurement location generates two adjacent timelike syndrome events.}
\label{fig:case3}
\end{figure}

\textit{Case 3.} The third process that contributes to the same order involves error chains weaving through surfaces in different time slices. In Fig.~\ref{fig:case3} we show a single row of a distance-5 surface in two subsequent time slices. Note that the geometry of locations of data qubits and measurement events for this case exactly correspond to the geometry of broken error chains discussed in case 2, except for the fact that the breaking point is along timelike direction instead of spacelike one. In analogy with case 2 we argue that such a situation happens for two subcases: ${\rm i\rangle}$ when there is one measurement error with a probability $q_{\rm bf}$ on one time slice along with $(d-1)/2$ bit-flip errors on data qubits in two subsequent time slices in a single row, and ${\rm ii\rangle}$ when there are only $(d+1)/2$ bit-flip errors on data qubits in two subsequent time slices in a single row. Bit-flip error probability on a syndrome or data qubit in one of the two subsequent time slices is in fact $p_{\rm bf}(1-p_{\rm bf})$ or $q_{\rm bf}(1-q_{\rm bf})$ and keeping only leading order terms we approximate those as $p_{\rm bf}$ or $q_{\rm bf}$. Note that the two subcases of case-3 can be mapped exactly with the two subcases of case-2 as far as their combinatorics are concerned and following a similar argument as in case 2 we obtain
\BEq
\label{eq:pxl3}
P_{\rm XL}^{(3)}=d\left[\frac{d^{2}-1}{d+3}\frac{q_{\rm bf}}{p_{\rm bf}}+\frac{d-1}{2}\right]{d \choose \frac{d-1}{2}}p_{\rm bf}^{\frac{d+1}{2}},
\EEq
where the difference in the prefactor comes from the fact that the single row for this case can be chosen in $d$ ways. Assuming $q_{\rm bf}$ is of the same order of magnitude as $p_{\rm bf}$, we observe that case 3 in fact contributes to the same order like previous cases. Also assuming $p_{\rm bf}=q_{\rm bf}$ and replacing the prefactor $d$ with $d-1$ in (\ref{eq:pxl3}), we can retrieve the right hand side of (\ref{eq:pxl2}). We claim that---except for these three cases---all other processes contribute higher-order terms as they involve multiple breaking points. Combining all the contributions we obtain
\BW
\BEq
\label{eq:pxl}
P_{\rm XL}=\left[d+(d-1)\frac{(3d+5)(d-1)}{2(d+3)}+d\left(\frac{d^{2}-1}{d+3}\frac{q_{\rm bf}}{p_{\rm bf}}+\frac{d-1}{2}\right)\right]{d \choose \frac{d-1}{2}}p_{\rm bf}^{\frac{d+1}{2}}.
\EEq
As far as the topology of the logical error chains are concerned, there is no difference between logical $X$ and $Z$ errors which enables us to use the same combinatorics to show that the logical $Z$ error probability,
\BEq
\label{eq:pzl}
P_{\rm ZL}=\left[d+(d-1)\frac{(3d+5)(d-1)}{2(d+3)}+d\left(\frac{d^{2}-1}{d+3}\frac{q_{\rm pf}}{p_{\rm pf}}+\frac{d-1}{2}\right)\right]{d \choose \frac{d-1}{2}}p_{\rm pf}^{\frac{d+1}{2}},
\EEq
\EW
where $p_{\rm pf}$ and $q_{\rm pf}$ are phase-flip error probabilities (per cycle) in data and syndrome qubits respectively.

At this point, we emphasize that our derivation never invokes any particular assumption about internal steps of a surface code cycle and therefore is also valid in a situation where the capability of directly measuring three or four qubit Pauli operators is implicitly assumed. As pointed out in Ref.~\cite{Fowler2011QIC,Fowl12g}, for a surface code cycle where measurement of multi-qubit operators are replaced by a sequence of CNOT operations, additional error chains having pure diagonal links emerge. While these error chains also contribute to the leading order, the number of such error chains is negligible for low distances. To verify the performance of our analytic expression, we assume a symmetric depolarization channel error model for an 8-step surface code cycle (as described in Ref.~\cite{PhysRevA.86.032324}) and plot logical $X$ error rate per cycle as a function of single physical qubit error rate per timestep ($p_{\rm step}$), which is (approximately) related to $p_{\rm bf}$ via
\BEq
p_{\rm step}=\frac{3}{2}\left(\frac{p_{\rm bf}}{8}\right).
\EEq
Fig.~\ref{fig:logicalErrorFormulaComp} shows a comparison (for logical $X$ error) of our analytical estimate and a numerical Monte-Carlo simulation as obtained in Ref.~\cite{PhysRevA.86.032324}; it is evident that for low distances the analytic estimate correctly captures the dominant behavior of these error chains below threshold.

\section{Coupled qubit model under decoherence}
\label{app:sec2}

In this section we compute the fidelity loss during a controlled-$Z$ (CZ) gate for a coupled qubit model under amplitude and phase damping. Such a model is important for the estimation of total CNOT gate time as well as intrinsic errors for textbook architecture. Since we assume the couplers having infinite on-off ratio for this architecture, each pair of qubits gets decoupled from all other pairs for each intermediate step of error correction cycle and therefore each pair of coupled qubits can be treated separately. Both the qubits are assumed to have three levels and the Hamiltonian is given by,
\BW
\BEq
\label{eq:2qubithamiltonian}
H(t) = \begin{pmatrix}
 0 & 0 & 0 \\
 0 & \omega_{\rm 1}(t) & 0 \\
 0 & 0 & 2\omega_{\rm 1}(t)-\eta \\
 \end{pmatrix}_{\rm \! \! q_{1}} + \begin{pmatrix}
 0 & 0 & 0  \\
 0 & \omega_{\rm 2} & 0 \\
 0 & 0 & 2\omega_{\rm 2}-\eta \\
 \end{pmatrix}_{\rm \! \! q_{2}}+g \begin{pmatrix}
 0 & -i & 0\\
 i & 0 & -i\sqrt{2}\\
 0 & i\sqrt{2} & 0\\
 \end{pmatrix}_{\rm \! \! q_{1}} \otimes \begin{pmatrix}
 0 & -i & 0 \\
 i & 0 & -i\sqrt{2} \\
 0 & i\sqrt{2} & 0 \\
 \end{pmatrix}_{\rm \! \! q_{2}},
\EEq
where the suffix denotes qubit index, $g$ represents the coupling between the qubits and $\eta$ is the anharmonicity of the qubit. For a CZ operation we control the frequency of the first qubit ($\omega_{\rm 1}(t)$) with an error function pulse as described in Ref.~\cite{ghosh2012CZ} while the frequency of the second qubit is kept constant. The Kraus matrices for the amplitude damping channel of any three level quantum system are given by,
\BEq
\label{eq:2qubitkrausAD}
E^{\rm AD}_{\rm 1}=\begin{pmatrix}
 1 & 0 & 0 \\
 0 & \sqrt{1-\lambda_{\rm 1}} & 0 \\
 0 & 0 & \sqrt{1-\lambda_{\rm 2}} \\
 \end{pmatrix}, \; E^{\rm AD}_{\rm 2}=\begin{pmatrix}
 0 & \sqrt{\lambda_{\rm 1}} & 0 \\
 0 & 0 & 0 \\
 0 & 0 & 0 \\
 \end{pmatrix}, \; E^{\rm AD}_{\rm 3}=\begin{pmatrix}
 0 & 0 & \sqrt{\lambda_{\rm 2}} \\
 0 & 0 & 0 \\
 0 & 0 & 0 \\
 \end{pmatrix}
\EEq
and Kraus matrices for phase damping are given by,
\BEq
\label{eq:2qubitkrausPD}
E^{\rm PD}_{\rm 1}=\begin{pmatrix}
 1 & 0 & 0 \\
 0 & \sqrt{1-\lambda_{\rm 3}} & 0 \\
 0 & 0 & \sqrt{1-\lambda_{\rm 4}} \\
 \end{pmatrix}, \; E^{\rm PD}_{\rm 2}=\begin{pmatrix}
 0 & 0 & 0 \\
 0 & \sqrt{\lambda_{\rm 3}} & 0 \\
 0 & 0 & \sqrt{\lambda_{\rm 4}} \\
 \end{pmatrix},
\EEq
\EW
where $\lambda_{k}$ for $k=1,2,3,4$ being parameters of our decoherence model. We assume the same amplitude and phase damping probability for $\ket{1}$ and $\ket{2}$ states ($\lambda \equiv \lambda_{\rm 1}=\lambda_{\rm 2}$ and $\lambda' \equiv \lambda_{\rm 3}=\lambda_{\rm 4}$) and represent $\lambda$ and $\lambda'$ as functions of time duration (${\Delta}t$) and $T_{\rm 1}$, $T_{\rm 2}$ of the quantum system as,
\BEq
\lambda({\Delta}t,T_{\rm 1})=1-e^{-{\Delta}t/T_{\rm 1}}, \lambda'({\Delta}t,T_{\rm 1},T_{\rm 2})=1-e^{-{\Delta}t\left[\frac{2}{T_{\rm 2}}-\frac{1}{T_{\rm 1}}\right]}.
\EEq

\begin{table}[htb]
\caption{Optimal parameters and results obtained for CNOT gate in this coupled qubit model. We use these results for the estimation of logical error rate in textbook architecture.}
\begin{tabular}{|c|c|c|c|c|c|}
\hline $\omega_{\rm 1}(t=0)$ & $\omega_{\rm 2}$ & $\eta$ & $g$ & $t_{\rm CNOT}$ & $p_{\rm intr}$ \\ 
\hline 8 GHz & 6 GHz & 300 MHz & 55 MHz & 21 ns & 1.23 $\times 10^{-4}$ \\ 
\hline 
\end{tabular}
\label{table:twoQubitModel}
\end{table}

\begin{figure}[htb]
\includegraphics[angle=0,width=\linewidth]{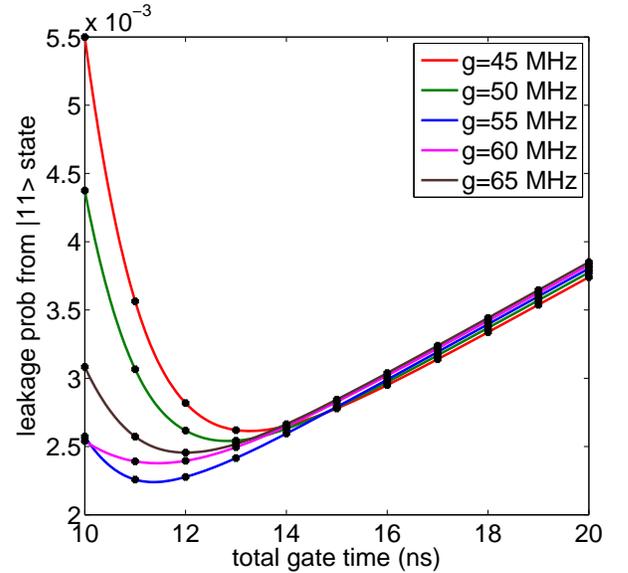}
\caption{(Color online) Plot of leakage probability from $\ket{11}$ state under decoherence for various $g$ during a CZ operation vs total CZ operation time.}
\label{fig:qresDecoh}
\end{figure}

The assumption that decoherence affects each qubit independently enables us to construct the full Kraus matrices for the qubit-qubit model by performing all possible tensor products between individual single qubit Kraus matrices. We first simulate the Hamiltonian given by (\ref{eq:2qubithamiltonian}) for parameters given in Table.~\ref{table:model} without decoherence to obtain an optimal pulse shape that maximizes the average fidelity of the CZ gate for a given coupling and gate time. Next we apply our decoherence model described by Eqs.~(\ref{eq:2qubitkrausAD}) and (\ref{eq:2qubitkrausPD}) on those optimal pulses. Fig.~\ref{fig:qresDecoh} shows plots of leakage error from $\ket{11}$ for such decoherence model (for $T_{1}=10~{\mu}{\rm s}$) applied on optimal pulses with respect to various total gate time and for various values of coupling strengths. Fig.~\ref{fig:qresDecoh} also shows that there exists an optimal point corresponding to total gate time $\sim$ 11 ns at $g=55$ MHz for which the leakage from $\ket{11}$ state is the minimum under decoherence. We use this point for the CZ part of the CNOT operation in textbook architecture and assuming that local rotations can be performed almost exactly in 5 ns, a CNOT requires 21 ns time duration as it involves two Hadamard operations along with a CZ. Table~\ref{table:twoQubitModel} shows the optimal parameters and results obtained from this analysis.

\bibliography{surfaceCodeArchitecture}

\begin{thebibliography}{41}%
\makeatletter
\providecommand \@ifxundefined [1]{%
 \@ifx{#1\undefined}
}%
\providecommand \@ifnum [1]{%
 \ifnum #1\expandafter \@firstoftwo
 \else \expandafter \@secondoftwo
 \fi
}%
\providecommand \@ifx [1]{%
 \ifx #1\expandafter \@firstoftwo
 \else \expandafter \@secondoftwo
 \fi
}%
\providecommand \natexlab [1]{#1}%
\providecommand \enquote  [1]{``#1''}%
\providecommand \bibnamefont  [1]{#1}%
\providecommand \bibfnamefont [1]{#1}%
\providecommand \citenamefont [1]{#1}%
\providecommand \href@noop [0]{\@secondoftwo}%
\providecommand \href [0]{\begingroup \@sanitize@url \@href}%
\providecommand \@href[1]{\@@startlink{#1}\@@href}%
\providecommand \@@href[1]{\endgroup#1\@@endlink}%
\providecommand \@sanitize@url [0]{\catcode `\\12\catcode `\$12\catcode
  `\&12\catcode `\#12\catcode `\^12\catcode `\_12\catcode `\%12\relax}%
\providecommand \@@startlink[1]{}%
\providecommand \@@endlink[0]{}%
\providecommand \url  [0]{\begingroup\@sanitize@url \@url }%
\providecommand \@url [1]{\endgroup\@href {#1}{\urlprefix }}%
\providecommand \urlprefix  [0]{URL }%
\providecommand \Eprint [0]{\href }%
\@ifxundefined \urlstyle {%
  \providecommand \doi  [0]{\begingroup \@sanitize@url \@doi}%
  \providecommand \@doi [1]{\endgroup \@@startlink {\doibase
  #1}doi:\discretionary {}{}{}#1\@@endlink }%
}{%
  \providecommand \doi  [0]{doi:\discretionary{}{}{}\begingroup
  \urlstyle{rm}\Url }%
}%
\providecommand \doibase [0]{http://dx.doi.org/}%
\providecommand \Doi [0]{\begingroup \@sanitize@url \@Doi }%
\providecommand \@Doi  [1]{\endgroup\@@startlink{\doibase#1}\@@Doi}%
\providecommand \@@Doi [1]{#1\@@endlink}%
\providecommand \selectlanguage [0]{\@gobble}%
\providecommand \bibinfo  [0]{\@secondoftwo}%
\providecommand \bibfield  [0]{\@secondoftwo}%
\providecommand \translation [1]{[#1]}%
\providecommand \BibitemOpen [0]{}%
\providecommand \bibitemStop [0]{}%
\providecommand \bibitemNoStop [0]{.\EOS\space}%
\providecommand \EOS [0]{\spacefactor3000\relax}%
\providecommand \BibitemShut  [1]{\csname bibitem#1\endcsname}%
\bibitem [{\citenamefont {Bravyi}\ and\ \citenamefont
  {Kitaev}(1998)}]{1998quant.ph.11052B}%
  \BibitemOpen
  \bibfield  {author} {\bibinfo {author} {\bibfnamefont {S.~B.}\ \bibnamefont
  {Bravyi}}\ and\ \bibinfo {author} {\bibfnamefont {A.~Y.}\ \bibnamefont
  {Kitaev}},\ }\href@noop {} { (\bibinfo {year} {1998})},\ \Eprint
  {http://arxiv.org/abs/1206.5407} {arXiv:1206.5407 [quant-ph]} \BibitemShut
  {NoStop}%
\bibitem [{\citenamefont {Dennis}\ \emph {et~al.}(2002)\citenamefont {Dennis},
  \citenamefont {Kitaev}, \citenamefont {Landahl},\ and\ \citenamefont
  {Preskill}}]{dennis:4452}%
  \BibitemOpen
  \bibfield  {author} {\bibinfo {author} {\bibfnamefont {E.}~\bibnamefont
  {Dennis}}, \bibinfo {author} {\bibfnamefont {A.}~\bibnamefont {Kitaev}},
  \bibinfo {author} {\bibfnamefont {A.}~\bibnamefont {Landahl}}, \ and\
  \bibinfo {author} {\bibfnamefont {J.}~\bibnamefont {Preskill}},\ }\Doi
  {10.1063/1.1499754} {\bibfield  {journal} {\bibinfo  {journal} {Journal of
  Mathematical Physics},\ }\textbf {\bibinfo {volume} {43}},\ \bibinfo {pages}
  {4452} (\bibinfo {year} {2002})}\BibitemShut {NoStop}%
\bibitem [{\citenamefont {Raussendorf}\ and\ \citenamefont
  {Harrington}(2007)}]{PhysRevLett.98.190504}%
  \BibitemOpen
  \bibfield  {author} {\bibinfo {author} {\bibfnamefont {R.}~\bibnamefont
  {Raussendorf}}\ and\ \bibinfo {author} {\bibfnamefont {J.}~\bibnamefont
  {Harrington}},\ }\Doi {10.1103/PhysRevLett.98.190504} {\bibfield  {journal}
  {\bibinfo  {journal} {Phys. Rev. Lett.},\ }\textbf {\bibinfo {volume} {98}},\
  \bibinfo {pages} {190504} (\bibinfo {year} {2007})}\BibitemShut {NoStop}%
\bibitem [{\citenamefont {Raussendorf}\ \emph {et~al.}(2007)\citenamefont
  {Raussendorf}, \citenamefont {Harrington},\ and\ \citenamefont
  {Goyal}}]{1367-2630-9-6-199}%
  \BibitemOpen
  \bibfield  {author} {\bibinfo {author} {\bibfnamefont {R.}~\bibnamefont
  {Raussendorf}}, \bibinfo {author} {\bibfnamefont {J.}~\bibnamefont
  {Harrington}}, \ and\ \bibinfo {author} {\bibfnamefont {K.}~\bibnamefont
  {Goyal}},\ }\href {http://stacks.iop.org/1367-2630/9/i=6/a=199} {\bibfield
  {journal} {\bibinfo  {journal} {New Journal of Physics},\ }\textbf {\bibinfo
  {volume} {9}},\ \bibinfo {pages} {199} (\bibinfo {year} {2007})}\BibitemShut
  {NoStop}%
\bibitem [{\citenamefont {Wang}\ \emph {et~al.}(2011)\citenamefont {Wang},
  \citenamefont {Fowler},\ and\ \citenamefont {Hollenberg}}]{Wang11}%
  \BibitemOpen
  \bibfield  {author} {\bibinfo {author} {\bibfnamefont {D.~S.}\ \bibnamefont
  {Wang}}, \bibinfo {author} {\bibfnamefont {A.~G.}\ \bibnamefont {Fowler}}, \
  and\ \bibinfo {author} {\bibfnamefont {L.~C.~L.}\ \bibnamefont
  {Hollenberg}},\ }\Doi {10.1103/PhysRevA.83.020302} {\bibfield  {journal}
  {\bibinfo  {journal} {Phys. Rev. A},\ }\textbf {\bibinfo {volume} {83}},\
  \bibinfo {pages} {020302} (\bibinfo {year} {2011})}\BibitemShut {NoStop}%
\bibitem [{\citenamefont {Fowler}\ \emph
  {et~al.}(2012){\natexlab{a}}\citenamefont {Fowler}, \citenamefont
  {Whiteside},\ and\ \citenamefont {Hollenberg}}]{Fowl11b}%
  \BibitemOpen
  \bibfield  {author} {\bibinfo {author} {\bibfnamefont {A.~G.}\ \bibnamefont
  {Fowler}}, \bibinfo {author} {\bibfnamefont {A.~C.}\ \bibnamefont
  {Whiteside}}, \ and\ \bibinfo {author} {\bibfnamefont {L.~C.~L.}\
  \bibnamefont {Hollenberg}},\ }\Doi {10.1103/PhysRevLett.108.180501}
  {\bibfield  {journal} {\bibinfo  {journal} {Phys. Rev. Lett.},\ }\textbf
  {\bibinfo {volume} {108}},\ \bibinfo {pages} {180501} (\bibinfo {year}
  {2012}{\natexlab{a}})}\BibitemShut {NoStop}%
\bibitem [{\citenamefont {Fowler}\ \emph
  {et~al.}(2012){\natexlab{b}}\citenamefont {Fowler}, \citenamefont
  {Mariantoni}, \citenamefont {Martinis},\ and\ \citenamefont
  {Cleland}}]{PhysRevA.86.032324}%
  \BibitemOpen
  \bibfield  {author} {\bibinfo {author} {\bibfnamefont {A.~G.}\ \bibnamefont
  {Fowler}}, \bibinfo {author} {\bibfnamefont {M.}~\bibnamefont {Mariantoni}},
  \bibinfo {author} {\bibfnamefont {J.~M.}\ \bibnamefont {Martinis}}, \ and\
  \bibinfo {author} {\bibfnamefont {A.~N.}\ \bibnamefont {Cleland}},\ }\Doi
  {10.1103/PhysRevA.86.032324} {\bibfield  {journal} {\bibinfo  {journal}
  {Phys. Rev. A},\ }\textbf {\bibinfo {volume} {86}},\ \bibinfo {pages}
  {032324} (\bibinfo {year} {2012}{\natexlab{b}})}\BibitemShut {NoStop}%
\bibitem [{\citenamefont {Koch}\ \emph {et~al.}(2007)\citenamefont {Koch},
  \citenamefont {Yu}, \citenamefont {Gambetta}, \citenamefont {Houck},
  \citenamefont {Schuster}, \citenamefont {Majer}, \citenamefont {Blais},
  \citenamefont {Devoret}, \citenamefont {Girvin},\ and\ \citenamefont
  {Schoelkopf}}]{PhysRevA.76.042319}%
  \BibitemOpen
  \bibfield  {author} {\bibinfo {author} {\bibfnamefont {J.}~\bibnamefont
  {Koch}}, \bibinfo {author} {\bibfnamefont {T.~M.}\ \bibnamefont {Yu}},
  \bibinfo {author} {\bibfnamefont {J.}~\bibnamefont {Gambetta}}, \bibinfo
  {author} {\bibfnamefont {A.~A.}\ \bibnamefont {Houck}}, \bibinfo {author}
  {\bibfnamefont {D.~I.}\ \bibnamefont {Schuster}}, \bibinfo {author}
  {\bibfnamefont {J.}~\bibnamefont {Majer}}, \bibinfo {author} {\bibfnamefont
  {A.}~\bibnamefont {Blais}}, \bibinfo {author} {\bibfnamefont {M.~H.}\
  \bibnamefont {Devoret}}, \bibinfo {author} {\bibfnamefont {S.~M.}\
  \bibnamefont {Girvin}}, \ and\ \bibinfo {author} {\bibfnamefont {R.~J.}\
  \bibnamefont {Schoelkopf}},\ }\Doi {10.1103/PhysRevA.76.042319} {\bibfield
  {journal} {\bibinfo  {journal} {Phys. Rev. A},\ }\textbf {\bibinfo {volume}
  {76}},\ \bibinfo {pages} {042319} (\bibinfo {year} {2007})}\BibitemShut
  {NoStop}%
\bibitem [{\citenamefont {Schreier}\ \emph {et~al.}(2008)\citenamefont
  {Schreier}, \citenamefont {Houck}, \citenamefont {Koch}, \citenamefont
  {Schuster}, \citenamefont {Johnson}, \citenamefont {Chow}, \citenamefont
  {Gambetta}, \citenamefont {Majer}, \citenamefont {Frunzio}, \citenamefont
  {Devoret}, \citenamefont {Girvin},\ and\ \citenamefont
  {Schoelkopf}}]{PhysRevB.77.180502}%
  \BibitemOpen
  \bibfield  {author} {\bibinfo {author} {\bibfnamefont {J.~A.}\ \bibnamefont
  {Schreier}}, \bibinfo {author} {\bibfnamefont {A.~A.}\ \bibnamefont {Houck}},
  \bibinfo {author} {\bibfnamefont {J.}~\bibnamefont {Koch}}, \bibinfo {author}
  {\bibfnamefont {D.~I.}\ \bibnamefont {Schuster}}, \bibinfo {author}
  {\bibfnamefont {B.~R.}\ \bibnamefont {Johnson}}, \bibinfo {author}
  {\bibfnamefont {J.~M.}\ \bibnamefont {Chow}}, \bibinfo {author}
  {\bibfnamefont {J.~M.}\ \bibnamefont {Gambetta}}, \bibinfo {author}
  {\bibfnamefont {J.}~\bibnamefont {Majer}}, \bibinfo {author} {\bibfnamefont
  {L.}~\bibnamefont {Frunzio}}, \bibinfo {author} {\bibfnamefont {M.~H.}\
  \bibnamefont {Devoret}}, \bibinfo {author} {\bibfnamefont {S.~M.}\
  \bibnamefont {Girvin}}, \ and\ \bibinfo {author} {\bibfnamefont {R.~J.}\
  \bibnamefont {Schoelkopf}},\ }\Doi {10.1103/PhysRevB.77.180502} {\bibfield
  {journal} {\bibinfo  {journal} {Phys. Rev. B},\ }\textbf {\bibinfo {volume}
  {77}},\ \bibinfo {pages} {180502} (\bibinfo {year} {2008})}\BibitemShut
  {NoStop}%
\bibitem [{\citenamefont {Paik}\ \emph {et~al.}(2011)\citenamefont {Paik},
  \citenamefont {Schuster}, \citenamefont {Bishop}, \citenamefont {Kirchmair},
  \citenamefont {Catelani}, \citenamefont {Sears}, \citenamefont {Johnson},
  \citenamefont {Reagor}, \citenamefont {Frunzio}, \citenamefont {Glazman},
  \citenamefont {Girvin}, \citenamefont {Devoret},\ and\ \citenamefont
  {Schoelkopf}}]{PhysRevLett.107.240501}%
  \BibitemOpen
  \bibfield  {author} {\bibinfo {author} {\bibfnamefont {H.}~\bibnamefont
  {Paik}}, \bibinfo {author} {\bibfnamefont {D.~I.}\ \bibnamefont {Schuster}},
  \bibinfo {author} {\bibfnamefont {L.~S.}\ \bibnamefont {Bishop}}, \bibinfo
  {author} {\bibfnamefont {G.}~\bibnamefont {Kirchmair}}, \bibinfo {author}
  {\bibfnamefont {G.}~\bibnamefont {Catelani}}, \bibinfo {author}
  {\bibfnamefont {A.~P.}\ \bibnamefont {Sears}}, \bibinfo {author}
  {\bibfnamefont {B.~R.}\ \bibnamefont {Johnson}}, \bibinfo {author}
  {\bibfnamefont {M.~J.}\ \bibnamefont {Reagor}}, \bibinfo {author}
  {\bibfnamefont {L.}~\bibnamefont {Frunzio}}, \bibinfo {author} {\bibfnamefont
  {L.~I.}\ \bibnamefont {Glazman}}, \bibinfo {author} {\bibfnamefont {S.~M.}\
  \bibnamefont {Girvin}}, \bibinfo {author} {\bibfnamefont {M.~H.}\
  \bibnamefont {Devoret}}, \ and\ \bibinfo {author} {\bibfnamefont {R.~J.}\
  \bibnamefont {Schoelkopf}},\ }\Doi {10.1103/PhysRevLett.107.240501}
  {\bibfield  {journal} {\bibinfo  {journal} {Phys. Rev. Lett.},\ }\textbf
  {\bibinfo {volume} {107}},\ \bibinfo {pages} {240501} (\bibinfo {year}
  {2011})}\BibitemShut {NoStop}%
\bibitem [{\citenamefont {Rigetti}\ \emph {et~al.}(2012)\citenamefont
  {Rigetti}, \citenamefont {Gambetta}, \citenamefont {Poletto}, \citenamefont
  {Plourde}, \citenamefont {Chow}, \citenamefont {C\'orcoles}, \citenamefont
  {Smolin}, \citenamefont {Merkel}, \citenamefont {Rozen}, \citenamefont
  {Keefe}, \citenamefont {Rothwell}, \citenamefont {Ketchen},\ and\
  \citenamefont {Steffen}}]{PhysRevB.86.100506}%
  \BibitemOpen
  \bibfield  {author} {\bibinfo {author} {\bibfnamefont {C.}~\bibnamefont
  {Rigetti}}, \bibinfo {author} {\bibfnamefont {J.~M.}\ \bibnamefont
  {Gambetta}}, \bibinfo {author} {\bibfnamefont {S.}~\bibnamefont {Poletto}},
  \bibinfo {author} {\bibfnamefont {B.~L.~T.}\ \bibnamefont {Plourde}},
  \bibinfo {author} {\bibfnamefont {J.~M.}\ \bibnamefont {Chow}}, \bibinfo
  {author} {\bibfnamefont {A.~D.}\ \bibnamefont {C\'orcoles}}, \bibinfo
  {author} {\bibfnamefont {J.~A.}\ \bibnamefont {Smolin}}, \bibinfo {author}
  {\bibfnamefont {S.~T.}\ \bibnamefont {Merkel}}, \bibinfo {author}
  {\bibfnamefont {J.~R.}\ \bibnamefont {Rozen}}, \bibinfo {author}
  {\bibfnamefont {G.~A.}\ \bibnamefont {Keefe}}, \bibinfo {author}
  {\bibfnamefont {M.~B.}\ \bibnamefont {Rothwell}}, \bibinfo {author}
  {\bibfnamefont {M.~B.}\ \bibnamefont {Ketchen}}, \ and\ \bibinfo {author}
  {\bibfnamefont {M.}~\bibnamefont {Steffen}},\ }\Doi
  {10.1103/PhysRevB.86.100506} {\bibfield  {journal} {\bibinfo  {journal}
  {Phys. Rev. B},\ }\textbf {\bibinfo {volume} {86}},\ \bibinfo {pages}
  {100506} (\bibinfo {year} {2012})}\BibitemShut {NoStop}%
\bibitem [{\citenamefont {Hime}\ \emph {et~al.}(2006)\citenamefont {Hime},
  \citenamefont {Reichardt}, \citenamefont {Plourde}, \citenamefont
  {Robertson}, \citenamefont {Wu}, \citenamefont {Ustinov},\ and\ \citenamefont
  {Clarke}}]{Hime01122006}%
  \BibitemOpen
  \bibfield  {author} {\bibinfo {author} {\bibfnamefont {T.}~\bibnamefont
  {Hime}}, \bibinfo {author} {\bibfnamefont {P.~A.}\ \bibnamefont {Reichardt}},
  \bibinfo {author} {\bibfnamefont {B.~L.~T.}\ \bibnamefont {Plourde}},
  \bibinfo {author} {\bibfnamefont {T.~L.}\ \bibnamefont {Robertson}}, \bibinfo
  {author} {\bibfnamefont {C.-E.}\ \bibnamefont {Wu}}, \bibinfo {author}
  {\bibfnamefont {A.~V.}\ \bibnamefont {Ustinov}}, \ and\ \bibinfo {author}
  {\bibfnamefont {J.}~\bibnamefont {Clarke}},\ }\Doi {10.1126/science.1134388}
  {\bibfield  {journal} {\bibinfo  {journal} {Science},\ }\textbf {\bibinfo
  {volume} {314}},\ \bibinfo {pages} {1427} (\bibinfo {year}
  {2006})}\BibitemShut {NoStop}%
\bibitem [{\citenamefont {van~der Ploeg}\ \emph {et~al.}(2007)\citenamefont
  {van~der Ploeg}, \citenamefont {Izmalkov}, \citenamefont {van~den Brink},
  \citenamefont {H\"ubner}, \citenamefont {Grajcar}, \citenamefont {Il'ichev},
  \citenamefont {Meyer},\ and\ \citenamefont
  {Zagoskin}}]{PhysRevLett.98.057004}%
  \BibitemOpen
  \bibfield  {author} {\bibinfo {author} {\bibfnamefont {S.~H.~W.}\
  \bibnamefont {van~der Ploeg}}, \bibinfo {author} {\bibfnamefont
  {A.}~\bibnamefont {Izmalkov}}, \bibinfo {author} {\bibfnamefont {A.~M.}\
  \bibnamefont {van~den Brink}}, \bibinfo {author} {\bibfnamefont
  {U.}~\bibnamefont {H\"ubner}}, \bibinfo {author} {\bibfnamefont
  {M.}~\bibnamefont {Grajcar}}, \bibinfo {author} {\bibfnamefont
  {E.}~\bibnamefont {Il'ichev}}, \bibinfo {author} {\bibfnamefont {H.-G.}\
  \bibnamefont {Meyer}}, \ and\ \bibinfo {author} {\bibfnamefont {A.~M.}\
  \bibnamefont {Zagoskin}},\ }\Doi {10.1103/PhysRevLett.98.057004} {\bibfield
  {journal} {\bibinfo  {journal} {Phys. Rev. Lett.},\ }\textbf {\bibinfo
  {volume} {98}},\ \bibinfo {pages} {057004} (\bibinfo {year}
  {2007})}\BibitemShut {NoStop}%
\bibitem [{\citenamefont {Niskanen}\ \emph {et~al.}(2007)\citenamefont
  {Niskanen}, \citenamefont {Harrabi}, \citenamefont {Yoshihara}, \citenamefont
  {Nakamura}, \citenamefont {Lloyd},\ and\ \citenamefont
  {Tsai}}]{Niskanen04052007}%
  \BibitemOpen
  \bibfield  {author} {\bibinfo {author} {\bibfnamefont {A.~O.}\ \bibnamefont
  {Niskanen}}, \bibinfo {author} {\bibfnamefont {K.}~\bibnamefont {Harrabi}},
  \bibinfo {author} {\bibfnamefont {F.}~\bibnamefont {Yoshihara}}, \bibinfo
  {author} {\bibfnamefont {Y.}~\bibnamefont {Nakamura}}, \bibinfo {author}
  {\bibfnamefont {S.}~\bibnamefont {Lloyd}}, \ and\ \bibinfo {author}
  {\bibfnamefont {J.~S.}\ \bibnamefont {Tsai}},\ }\Doi
  {10.1126/science.1141324} {\bibfield  {journal} {\bibinfo  {journal}
  {Science},\ }\textbf {\bibinfo {volume} {316}},\ \bibinfo {pages} {723}
  (\bibinfo {year} {2007})}\BibitemShut {NoStop}%
\bibitem [{\citenamefont {Harris}\ \emph {et~al.}(2007)\citenamefont {Harris},
  \citenamefont {Berkley}, \citenamefont {Johnson}, \citenamefont {Bunyk},
  \citenamefont {Govorkov}, \citenamefont {Thom}, \citenamefont {Uchaikin},
  \citenamefont {Wilson}, \citenamefont {Chung}, \citenamefont {Holtham},
  \citenamefont {Biamonte}, \citenamefont {Smirnov}, \citenamefont {Amin},\
  and\ \citenamefont {Maassen van~den Brink}}]{PhysRevLett.98.177001}%
  \BibitemOpen
  \bibfield  {author} {\bibinfo {author} {\bibfnamefont {R.}~\bibnamefont
  {Harris}}, \bibinfo {author} {\bibfnamefont {A.~J.}\ \bibnamefont {Berkley}},
  \bibinfo {author} {\bibfnamefont {M.~W.}\ \bibnamefont {Johnson}}, \bibinfo
  {author} {\bibfnamefont {P.}~\bibnamefont {Bunyk}}, \bibinfo {author}
  {\bibfnamefont {S.}~\bibnamefont {Govorkov}}, \bibinfo {author}
  {\bibfnamefont {M.~C.}\ \bibnamefont {Thom}}, \bibinfo {author}
  {\bibfnamefont {S.}~\bibnamefont {Uchaikin}}, \bibinfo {author}
  {\bibfnamefont {A.~B.}\ \bibnamefont {Wilson}}, \bibinfo {author}
  {\bibfnamefont {J.}~\bibnamefont {Chung}}, \bibinfo {author} {\bibfnamefont
  {E.}~\bibnamefont {Holtham}}, \bibinfo {author} {\bibfnamefont {J.~D.}\
  \bibnamefont {Biamonte}}, \bibinfo {author} {\bibfnamefont {A.~Y.}\
  \bibnamefont {Smirnov}}, \bibinfo {author} {\bibfnamefont {M.~H.~S.}\
  \bibnamefont {Amin}}, \ and\ \bibinfo {author} {\bibfnamefont
  {A.}~\bibnamefont {Maassen van~den Brink}},\ }\Doi
  {10.1103/PhysRevLett.98.177001} {\bibfield  {journal} {\bibinfo  {journal}
  {Phys. Rev. Lett.},\ }\textbf {\bibinfo {volume} {98}},\ \bibinfo {pages}
  {177001} (\bibinfo {year} {2007})}\BibitemShut {NoStop}%
\bibitem [{\citenamefont {Allman}\ \emph {et~al.}(2010)\citenamefont {Allman},
  \citenamefont {Altomare}, \citenamefont {Whittaker}, \citenamefont {Cicak},
  \citenamefont {Li}, \citenamefont {Sirois}, \citenamefont {Strong},
  \citenamefont {Teufel},\ and\ \citenamefont
  {Simmonds}}]{PhysRevLett.104.177004}%
  \BibitemOpen
  \bibfield  {author} {\bibinfo {author} {\bibfnamefont {M.~S.}\ \bibnamefont
  {Allman}}, \bibinfo {author} {\bibfnamefont {F.}~\bibnamefont {Altomare}},
  \bibinfo {author} {\bibfnamefont {J.~D.}\ \bibnamefont {Whittaker}}, \bibinfo
  {author} {\bibfnamefont {K.}~\bibnamefont {Cicak}}, \bibinfo {author}
  {\bibfnamefont {D.}~\bibnamefont {Li}}, \bibinfo {author} {\bibfnamefont
  {A.}~\bibnamefont {Sirois}}, \bibinfo {author} {\bibfnamefont
  {J.}~\bibnamefont {Strong}}, \bibinfo {author} {\bibfnamefont {J.~D.}\
  \bibnamefont {Teufel}}, \ and\ \bibinfo {author} {\bibfnamefont {R.~W.}\
  \bibnamefont {Simmonds}},\ }\Doi {10.1103/PhysRevLett.104.177004} {\bibfield
  {journal} {\bibinfo  {journal} {Phys. Rev. Lett.},\ }\textbf {\bibinfo
  {volume} {104}},\ \bibinfo {pages} {177004} (\bibinfo {year}
  {2010})}\BibitemShut {NoStop}%
\bibitem [{\citenamefont {Pinto}\ \emph {et~al.}(2010)\citenamefont {Pinto},
  \citenamefont {Korotkov}, \citenamefont {Geller}, \citenamefont {Shumeiko},\
  and\ \citenamefont {Martinis}}]{PhysRevB.82.104522}%
  \BibitemOpen
  \bibfield  {author} {\bibinfo {author} {\bibfnamefont {R.~A.}\ \bibnamefont
  {Pinto}}, \bibinfo {author} {\bibfnamefont {A.~N.}\ \bibnamefont {Korotkov}},
  \bibinfo {author} {\bibfnamefont {M.~R.}\ \bibnamefont {Geller}}, \bibinfo
  {author} {\bibfnamefont {V.~S.}\ \bibnamefont {Shumeiko}}, \ and\ \bibinfo
  {author} {\bibfnamefont {J.~M.}\ \bibnamefont {Martinis}},\ }\Doi
  {10.1103/PhysRevB.82.104522} {\bibfield  {journal} {\bibinfo  {journal}
  {Phys. Rev. B},\ }\textbf {\bibinfo {volume} {82}},\ \bibinfo {pages}
  {104522} (\bibinfo {year} {2010})}\BibitemShut {NoStop}%
\bibitem [{\citenamefont {Bialczak}\ \emph {et~al.}(2011)\citenamefont
  {Bialczak}, \citenamefont {Ansmann}, \citenamefont {Hofheinz}, \citenamefont
  {Lenander}, \citenamefont {Lucero}, \citenamefont {Neeley}, \citenamefont
  {O'Connell}, \citenamefont {Sank}, \citenamefont {Wang}, \citenamefont
  {Weides}, \citenamefont {Wenner}, \citenamefont {Yamamoto}, \citenamefont
  {Cleland},\ and\ \citenamefont {Martinis}}]{PhysRevLett.106.060501}%
  \BibitemOpen
  \bibfield  {author} {\bibinfo {author} {\bibfnamefont {R.~C.}\ \bibnamefont
  {Bialczak}}, \bibinfo {author} {\bibfnamefont {M.}~\bibnamefont {Ansmann}},
  \bibinfo {author} {\bibfnamefont {M.}~\bibnamefont {Hofheinz}}, \bibinfo
  {author} {\bibfnamefont {M.}~\bibnamefont {Lenander}}, \bibinfo {author}
  {\bibfnamefont {E.}~\bibnamefont {Lucero}}, \bibinfo {author} {\bibfnamefont
  {M.}~\bibnamefont {Neeley}}, \bibinfo {author} {\bibfnamefont {A.~D.}\
  \bibnamefont {O'Connell}}, \bibinfo {author} {\bibfnamefont {D.}~\bibnamefont
  {Sank}}, \bibinfo {author} {\bibfnamefont {H.}~\bibnamefont {Wang}}, \bibinfo
  {author} {\bibfnamefont {M.}~\bibnamefont {Weides}}, \bibinfo {author}
  {\bibfnamefont {J.}~\bibnamefont {Wenner}}, \bibinfo {author} {\bibfnamefont
  {T.}~\bibnamefont {Yamamoto}}, \bibinfo {author} {\bibfnamefont {A.~N.}\
  \bibnamefont {Cleland}}, \ and\ \bibinfo {author} {\bibfnamefont {J.~M.}\
  \bibnamefont {Martinis}},\ }\Doi {10.1103/PhysRevLett.106.060501} {\bibfield
  {journal} {\bibinfo  {journal} {Phys. Rev. Lett.},\ }\textbf {\bibinfo
  {volume} {106}},\ \bibinfo {pages} {060501} (\bibinfo {year}
  {2011})}\BibitemShut {NoStop}%
\bibitem [{\citenamefont {Groszkowski}\ \emph {et~al.}(2011)\citenamefont
  {Groszkowski}, \citenamefont {Fowler}, \citenamefont {Motzoi},\ and\
  \citenamefont {Wilhelm}}]{PhysRevB.84.144516}%
  \BibitemOpen
  \bibfield  {author} {\bibinfo {author} {\bibfnamefont {P.}~\bibnamefont
  {Groszkowski}}, \bibinfo {author} {\bibfnamefont {A.~G.}\ \bibnamefont
  {Fowler}}, \bibinfo {author} {\bibfnamefont {F.}~\bibnamefont {Motzoi}}, \
  and\ \bibinfo {author} {\bibfnamefont {F.~K.}\ \bibnamefont {Wilhelm}},\
  }\Doi {10.1103/PhysRevB.84.144516} {\bibfield  {journal} {\bibinfo  {journal}
  {Phys. Rev. B},\ }\textbf {\bibinfo {volume} {84}},\ \bibinfo {pages}
  {144516} (\bibinfo {year} {2011})}\BibitemShut {NoStop}%
\bibitem [{\citenamefont {DiCarlo}\ \emph {et~al.}(2009)\citenamefont
  {DiCarlo}, \citenamefont {Chow}, \citenamefont {Gambetta}, \citenamefont
  {Bishop}, \citenamefont {Johnson}, \citenamefont {Schuster}, \citenamefont
  {Majer}, \citenamefont {Blais}, \citenamefont {Frunzio}, \citenamefont
  {Girvin},\ and\ \citenamefont {Schoelkopf}}]{DiCarlo2009}%
  \BibitemOpen
  \bibfield  {author} {\bibinfo {author} {\bibfnamefont {L.}~\bibnamefont
  {DiCarlo}}, \bibinfo {author} {\bibfnamefont {J.~M.}\ \bibnamefont {Chow}},
  \bibinfo {author} {\bibfnamefont {J.~M.}\ \bibnamefont {Gambetta}}, \bibinfo
  {author} {\bibfnamefont {L.~S.}\ \bibnamefont {Bishop}}, \bibinfo {author}
  {\bibfnamefont {B.~R.}\ \bibnamefont {Johnson}}, \bibinfo {author}
  {\bibfnamefont {D.~I.}\ \bibnamefont {Schuster}}, \bibinfo {author}
  {\bibfnamefont {J.}~\bibnamefont {Majer}}, \bibinfo {author} {\bibfnamefont
  {A.}~\bibnamefont {Blais}}, \bibinfo {author} {\bibfnamefont
  {L.}~\bibnamefont {Frunzio}}, \bibinfo {author} {\bibfnamefont {S.~M.}\
  \bibnamefont {Girvin}}, \ and\ \bibinfo {author} {\bibfnamefont {R.~J.}\
  \bibnamefont {Schoelkopf}},\ }\href {http://dx.doi.org/10.1038/nature08121}
  {\bibfield  {journal} {\bibinfo  {journal} {Nature},\ }\textbf {\bibinfo
  {volume} {460}},\ \bibinfo {pages} {240} (\bibinfo {year}
  {2009})}\BibitemShut {NoStop}%
\bibitem [{\citenamefont {Strauch}\ \emph {et~al.}(2003)\citenamefont
  {Strauch}, \citenamefont {Johnson}, \citenamefont {Dragt}, \citenamefont
  {Lobb}, \citenamefont {Anderson},\ and\ \citenamefont
  {Wellstood}}]{StrauchPRL03}%
  \BibitemOpen
  \bibfield  {author} {\bibinfo {author} {\bibfnamefont {F.~W.}\ \bibnamefont
  {Strauch}}, \bibinfo {author} {\bibfnamefont {P.~R.}\ \bibnamefont
  {Johnson}}, \bibinfo {author} {\bibfnamefont {A.~J.}\ \bibnamefont {Dragt}},
  \bibinfo {author} {\bibfnamefont {C.~J.}\ \bibnamefont {Lobb}}, \bibinfo
  {author} {\bibfnamefont {J.~R.}\ \bibnamefont {Anderson}}, \ and\ \bibinfo
  {author} {\bibfnamefont {F.~C.}\ \bibnamefont {Wellstood}},\ }\href@noop {}
  {\bibfield  {journal} {\bibinfo  {journal} {Phys. Rev. Lett.},\ }\textbf
  {\bibinfo {volume} {91}},\ \bibinfo {pages} {167005} (\bibinfo {year}
  {2003})}\BibitemShut {NoStop}%
\bibitem [{\citenamefont {Ghosh}\ \emph {et~al.}()\citenamefont {Ghosh},
  \citenamefont {Galiautdinov}, \citenamefont {Zhou}, \citenamefont {Korotkov},
  \citenamefont {Martinis},\ and\ \citenamefont {Geller}}]{ghosh2012CZ}%
  \BibitemOpen
  \bibfield  {author} {\bibinfo {author} {\bibfnamefont {J.}~\bibnamefont
  {Ghosh}}, \bibinfo {author} {\bibfnamefont {A.}~\bibnamefont {Galiautdinov}},
  \bibinfo {author} {\bibfnamefont {Z.}~\bibnamefont {Zhou}}, \bibinfo {author}
  {\bibfnamefont {A.~N.}\ \bibnamefont {Korotkov}}, \bibinfo {author}
  {\bibfnamefont {J.~M.}\ \bibnamefont {Martinis}}, \ and\ \bibinfo {author}
  {\bibfnamefont {M.~R.}\ \bibnamefont {Geller}},\ }\href@noop {} {\bibinfo
  {journal} {``High-fidelity CZ gate for resonator-based superconducting
  quantum computers", unpublished}}\BibitemShut {NoStop}%
\bibitem [{\citenamefont {Motzoi}\ \emph {et~al.}(2009)\citenamefont {Motzoi},
  \citenamefont {Gambetta}, \citenamefont {Rebentrost},\ and\ \citenamefont
  {Wilhelm}}]{motzoi2009prl}%
  \BibitemOpen
\bibfield  {journal} {  }\bibfield  {author} {\bibinfo {author} {\bibfnamefont
  {F.}~\bibnamefont {Motzoi}}, \bibinfo {author} {\bibfnamefont {J.~M.}\
  \bibnamefont {Gambetta}}, \bibinfo {author} {\bibfnamefont {P.}~\bibnamefont
  {Rebentrost}}, \ and\ \bibinfo {author} {\bibfnamefont {F.~K.}\ \bibnamefont
  {Wilhelm}},\ }\Doi {10.1103/PhysRevLett.103.110501} {\bibfield  {journal}
  {\bibinfo  {journal} {Phys. Rev. Lett.},\ }\textbf {\bibinfo {volume}
  {103}},\ \bibinfo {pages} {110501} (\bibinfo {year} {2009})}\BibitemShut
  {NoStop}%
\bibitem [{Note1()}]{Note1}%
  \BibitemOpen
  \bibinfo {note} {E. A. Sete, A. Galiautdinov, E. Mlinar, J. M. Martinis, and
  A. N. Korotkov, unpublished.}\BibitemShut {Stop}%
\bibitem [{\citenamefont {Helmer}\ \emph {et~al.}(2009)\citenamefont {Helmer},
  \citenamefont {Mariantoni}, \citenamefont {Fowler}, \citenamefont {von
  Delft}, \citenamefont {Solano},\ and\ \citenamefont
  {Marquardt}}]{0295-5075-85-5-50007}%
  \BibitemOpen
  \bibfield  {author} {\bibinfo {author} {\bibfnamefont {F.}~\bibnamefont
  {Helmer}}, \bibinfo {author} {\bibfnamefont {M.}~\bibnamefont {Mariantoni}},
  \bibinfo {author} {\bibfnamefont {A.~G.}\ \bibnamefont {Fowler}}, \bibinfo
  {author} {\bibfnamefont {J.}~\bibnamefont {von Delft}}, \bibinfo {author}
  {\bibfnamefont {E.}~\bibnamefont {Solano}}, \ and\ \bibinfo {author}
  {\bibfnamefont {F.}~\bibnamefont {Marquardt}},\ }\href
  {http://stacks.iop.org/0295-5075/85/i=5/a=50007} {\bibfield  {journal}
  {\bibinfo  {journal} {EPL (Europhysics Letters)},\ }\textbf {\bibinfo
  {volume} {85}},\ \bibinfo {pages} {50007} (\bibinfo {year}
  {2009})}\BibitemShut {NoStop}%
\bibitem [{\citenamefont {DiVincenzo}(2009)}]{1402-4896-2009-T137-014020}%
  \BibitemOpen
  \bibfield  {author} {\bibinfo {author} {\bibfnamefont {D.~P.}\ \bibnamefont
  {DiVincenzo}},\ }\href {http://stacks.iop.org/1402-4896/2009/i=T137/a=014020}
  {\bibfield  {journal} {\bibinfo  {journal} {Physica Scripta},\ }\textbf
  {\bibinfo {volume} {T137}},\ \bibinfo {pages} {014020} (\bibinfo {year}
  {2009})}\BibitemShut {NoStop}%
\bibitem [{\citenamefont {Chow}\ \emph {et~al.}(2011)\citenamefont {Chow},
  \citenamefont {C\'orcoles}, \citenamefont {Gambetta}, \citenamefont
  {Rigetti}, \citenamefont {Johnson}, \citenamefont {Smolin}, \citenamefont
  {Rozen}, \citenamefont {Keefe}, \citenamefont {Rothwell}, \citenamefont
  {Ketchen},\ and\ \citenamefont {Steffen}}]{PhysRevLett.107.080502}%
  \BibitemOpen
  \bibfield  {author} {\bibinfo {author} {\bibfnamefont {J.~M.}\ \bibnamefont
  {Chow}}, \bibinfo {author} {\bibfnamefont {A.~D.}\ \bibnamefont
  {C\'orcoles}}, \bibinfo {author} {\bibfnamefont {J.~M.}\ \bibnamefont
  {Gambetta}}, \bibinfo {author} {\bibfnamefont {C.}~\bibnamefont {Rigetti}},
  \bibinfo {author} {\bibfnamefont {B.~R.}\ \bibnamefont {Johnson}}, \bibinfo
  {author} {\bibfnamefont {J.~A.}\ \bibnamefont {Smolin}}, \bibinfo {author}
  {\bibfnamefont {J.~R.}\ \bibnamefont {Rozen}}, \bibinfo {author}
  {\bibfnamefont {G.~A.}\ \bibnamefont {Keefe}}, \bibinfo {author}
  {\bibfnamefont {M.~B.}\ \bibnamefont {Rothwell}}, \bibinfo {author}
  {\bibfnamefont {M.~B.}\ \bibnamefont {Ketchen}}, \ and\ \bibinfo {author}
  {\bibfnamefont {M.}~\bibnamefont {Steffen}},\ }\Doi
  {10.1103/PhysRevLett.107.080502} {\bibfield  {journal} {\bibinfo  {journal}
  {Phys. Rev. Lett.},\ }\textbf {\bibinfo {volume} {107}},\ \bibinfo {pages}
  {080502} (\bibinfo {year} {2011})}\BibitemShut {NoStop}%
\bibitem [{\citenamefont {Chow}\ \emph {et~al.}(2012)\citenamefont {Chow},
  \citenamefont {{Gambetta}}, \citenamefont {{Corcoles}}, \citenamefont
  {{Merkel}}, \citenamefont {{Smolin}}, \citenamefont {{Rigetti}},
  \citenamefont {{Poletto}}, \citenamefont {{Keefe}}, \citenamefont
  {{Rothwell}}, \citenamefont {{Rozen}}, \citenamefont {{Ketchen}},\ and\
  \citenamefont {{Steffen}}}]{2012arXiv1202.5344C}%
  \BibitemOpen
  \bibfield  {author} {\bibinfo {author} {\bibfnamefont {J.~M.}\ \bibnamefont
  {Chow}}, \bibinfo {author} {\bibfnamefont {J.~M.}\ \bibnamefont
  {{Gambetta}}}, \bibinfo {author} {\bibfnamefont {A.~D.}\ \bibnamefont
  {{Corcoles}}}, \bibinfo {author} {\bibfnamefont {S.~T.}\ \bibnamefont
  {{Merkel}}}, \bibinfo {author} {\bibfnamefont {J.~A.}\ \bibnamefont
  {{Smolin}}}, \bibinfo {author} {\bibfnamefont {C.}~\bibnamefont {{Rigetti}}},
  \bibinfo {author} {\bibfnamefont {S.}~\bibnamefont {{Poletto}}}, \bibinfo
  {author} {\bibfnamefont {G.~A.}\ \bibnamefont {{Keefe}}}, \bibinfo {author}
  {\bibfnamefont {M.~B.}\ \bibnamefont {{Rothwell}}}, \bibinfo {author}
  {\bibfnamefont {J.~R.}\ \bibnamefont {{Rozen}}}, \bibinfo {author}
  {\bibfnamefont {M.~B.}\ \bibnamefont {{Ketchen}}}, \ and\ \bibinfo {author}
  {\bibfnamefont {M.}~\bibnamefont {{Steffen}}},\ }\href@noop {} { (\bibinfo
  {year} {2012})},\ \Eprint {http://arxiv.org/abs/1202.5344} {arXiv:1202.5344
  [quant-ph]} \BibitemShut {NoStop}%
\bibitem [{\citenamefont {Rigetti}\ and\ \citenamefont
  {Devoret}(2010)}]{PhysRevB.81.134507}%
  \BibitemOpen
  \bibfield  {author} {\bibinfo {author} {\bibfnamefont {C.}~\bibnamefont
  {Rigetti}}\ and\ \bibinfo {author} {\bibfnamefont {M.}~\bibnamefont
  {Devoret}},\ }\Doi {10.1103/PhysRevB.81.134507} {\bibfield  {journal}
  {\bibinfo  {journal} {Phys. Rev. B},\ }\textbf {\bibinfo {volume} {81}},\
  \bibinfo {pages} {134507} (\bibinfo {year} {2010})}\BibitemShut {NoStop}%
\bibitem [{\citenamefont {D\"ur}\ \emph {et~al.}(2005)\citenamefont {D\"ur},
  \citenamefont {Hein}, \citenamefont {Cirac},\ and\ \citenamefont
  {Briegel}}]{PhysRevA.72.052326}%
  \BibitemOpen
  \bibfield  {author} {\bibinfo {author} {\bibfnamefont {W.}~\bibnamefont
  {D\"ur}}, \bibinfo {author} {\bibfnamefont {M.}~\bibnamefont {Hein}},
  \bibinfo {author} {\bibfnamefont {J.~I.}\ \bibnamefont {Cirac}}, \ and\
  \bibinfo {author} {\bibfnamefont {H.-J.}\ \bibnamefont {Briegel}},\ }\Doi
  {10.1103/PhysRevA.72.052326} {\bibfield  {journal} {\bibinfo  {journal}
  {Phys. Rev. A},\ }\textbf {\bibinfo {volume} {72}},\ \bibinfo {pages}
  {052326} (\bibinfo {year} {2005})}\BibitemShut {NoStop}%
\bibitem [{\citenamefont {Emerson}\ \emph {et~al.}(2007)\citenamefont
  {Emerson}, \citenamefont {Silva}, \citenamefont {Moussa}, \citenamefont
  {Ryan}, \citenamefont {Laforest}, \citenamefont {Baugh}, \citenamefont
  {Cory},\ and\ \citenamefont {Laflamme}}]{Emerson28092007}%
  \BibitemOpen
  \bibfield  {author} {\bibinfo {author} {\bibfnamefont {J.}~\bibnamefont
  {Emerson}}, \bibinfo {author} {\bibfnamefont {M.}~\bibnamefont {Silva}},
  \bibinfo {author} {\bibfnamefont {O.}~\bibnamefont {Moussa}}, \bibinfo
  {author} {\bibfnamefont {C.}~\bibnamefont {Ryan}}, \bibinfo {author}
  {\bibfnamefont {M.}~\bibnamefont {Laforest}}, \bibinfo {author}
  {\bibfnamefont {J.}~\bibnamefont {Baugh}}, \bibinfo {author} {\bibfnamefont
  {D.~G.}\ \bibnamefont {Cory}}, \ and\ \bibinfo {author} {\bibfnamefont
  {R.}~\bibnamefont {Laflamme}},\ }\Doi {10.1126/science.1145699} {\bibfield
  {journal} {\bibinfo  {journal} {Science},\ }\textbf {\bibinfo {volume}
  {317}},\ \bibinfo {pages} {1893} (\bibinfo {year} {2007})}\BibitemShut
  {NoStop}%
\bibitem [{\citenamefont {Silva}\ \emph {et~al.}(2008)\citenamefont {Silva},
  \citenamefont {Magesan}, \citenamefont {Kribs},\ and\ \citenamefont
  {Emerson}}]{PhysRevA.78.012347}%
  \BibitemOpen
  \bibfield  {author} {\bibinfo {author} {\bibfnamefont {M.}~\bibnamefont
  {Silva}}, \bibinfo {author} {\bibfnamefont {E.}~\bibnamefont {Magesan}},
  \bibinfo {author} {\bibfnamefont {D.~W.}\ \bibnamefont {Kribs}}, \ and\
  \bibinfo {author} {\bibfnamefont {J.}~\bibnamefont {Emerson}},\ }\Doi
  {10.1103/PhysRevA.78.012347} {\bibfield  {journal} {\bibinfo  {journal}
  {Phys. Rev. A},\ }\textbf {\bibinfo {volume} {78}},\ \bibinfo {pages}
  {012347} (\bibinfo {year} {2008})}\BibitemShut {NoStop}%
\bibitem [{\citenamefont {Dankert}\ \emph {et~al.}(2009)\citenamefont
  {Dankert}, \citenamefont {Cleve}, \citenamefont {Emerson},\ and\
  \citenamefont {Livine}}]{PhysRevA.80.012304}%
  \BibitemOpen
  \bibfield  {author} {\bibinfo {author} {\bibfnamefont {C.}~\bibnamefont
  {Dankert}}, \bibinfo {author} {\bibfnamefont {R.}~\bibnamefont {Cleve}},
  \bibinfo {author} {\bibfnamefont {J.}~\bibnamefont {Emerson}}, \ and\
  \bibinfo {author} {\bibfnamefont {E.}~\bibnamefont {Livine}},\ }\Doi
  {10.1103/PhysRevA.80.012304} {\bibfield  {journal} {\bibinfo  {journal}
  {Phys. Rev. A},\ }\textbf {\bibinfo {volume} {80}},\ \bibinfo {pages}
  {012304} (\bibinfo {year} {2009})}\BibitemShut {NoStop}%
\bibitem [{\citenamefont {Sarvepalli}\ \emph {et~al.}(2009)\citenamefont
  {Sarvepalli}, \citenamefont {Klappenecker},\ and\ \citenamefont
  {R\"{o}tteler}}]{Sarvepalli08052009}%
  \BibitemOpen
  \bibfield  {author} {\bibinfo {author} {\bibfnamefont {P.~K.}\ \bibnamefont
  {Sarvepalli}}, \bibinfo {author} {\bibfnamefont {A.}~\bibnamefont
  {Klappenecker}}, \ and\ \bibinfo {author} {\bibfnamefont {M.}~\bibnamefont
  {R\"{o}tteler}},\ }\Doi {10.1098/rspa.2008.0439} {\bibfield  {journal}
  {\bibinfo  {journal} {Proceedings of the Royal Society A: Mathematical,
  Physical and Engineering Science},\ }\textbf {\bibinfo {volume} {465}},\
  \bibinfo {pages} {1645} (\bibinfo {year} {2009})}\BibitemShut {NoStop}%
\bibitem [{\citenamefont {Magesan}\ \emph {et~al.}(2012)\citenamefont
  {Magesan}, \citenamefont {Puzzuoli}, \citenamefont {Granade},\ and\
  \citenamefont {Cory}}]{2012arXiv1206.5407M}%
  \BibitemOpen
  \bibfield  {author} {\bibinfo {author} {\bibfnamefont {E.}~\bibnamefont
  {Magesan}}, \bibinfo {author} {\bibfnamefont {D.}~\bibnamefont {Puzzuoli}},
  \bibinfo {author} {\bibfnamefont {C.~E.}\ \bibnamefont {Granade}}, \ and\
  \bibinfo {author} {\bibfnamefont {D.~G.}\ \bibnamefont {Cory}},\ }\href@noop
  {} { (\bibinfo {year} {2012})},\ \Eprint {http://arxiv.org/abs/1206.5407}
  {arXiv:1206.5407 [quant-ph]} \BibitemShut {NoStop}%
\bibitem [{\citenamefont {Guti{\'e}rrez}\ \emph {et~al.}(2012)\citenamefont
  {Guti{\'e}rrez}, \citenamefont {Svec}, \citenamefont {Vargo},\ and\
  \citenamefont {Brown}}]{2012arXiv1207.0046G}%
  \BibitemOpen
  \bibfield  {author} {\bibinfo {author} {\bibfnamefont {M.}~\bibnamefont
  {Guti{\'e}rrez}}, \bibinfo {author} {\bibfnamefont {L.}~\bibnamefont {Svec}},
  \bibinfo {author} {\bibfnamefont {A.}~\bibnamefont {Vargo}}, \ and\ \bibinfo
  {author} {\bibfnamefont {K.~R.}\ \bibnamefont {Brown}},\ }\href@noop {} {
  (\bibinfo {year} {2012})},\ \Eprint {http://arxiv.org/abs/1207.0046}
  {arXiv:1207.0046 [quant-ph]} \BibitemShut {NoStop}%
\bibitem [{\citenamefont {Wilhelm}\ \emph {et~al.}(2007)\citenamefont
  {Wilhelm}, \citenamefont {Storcz}, \citenamefont {Hartmann},\ and\
  \citenamefont {Geller}}]{Geller2007Springer2}%
  \BibitemOpen
  \bibfield  {author} {\bibinfo {author} {\bibfnamefont {F.~K.}\ \bibnamefont
  {Wilhelm}}, \bibinfo {author} {\bibfnamefont {M.~J.}\ \bibnamefont {Storcz}},
  \bibinfo {author} {\bibfnamefont {U.}~\bibnamefont {Hartmann}}, \ and\
  \bibinfo {author} {\bibfnamefont {M.~R.}\ \bibnamefont {Geller}},\
  }\href@noop {} {\bibfield  {journal} {\bibinfo  {journal} {Manipulating
  Quantum Coherence in Solid State Systems, edited by M. E. Flatte and I.
  Tifrea (Springer, 2007).},\ }\textbf {\bibinfo {volume} {Springer, 2007}},\
  \bibinfo {pages} {195} (\bibinfo {year} {2007})}\BibitemShut {NoStop}%
\bibitem [{\citenamefont {Galiautdinov}\ \emph {et~al.}(2012)\citenamefont
  {Galiautdinov}, \citenamefont {Korotkov},\ and\ \citenamefont
  {Martinis}}]{PhysRevA.85.042321}%
  \BibitemOpen
  \bibfield  {author} {\bibinfo {author} {\bibfnamefont {A.}~\bibnamefont
  {Galiautdinov}}, \bibinfo {author} {\bibfnamefont {A.~N.}\ \bibnamefont
  {Korotkov}}, \ and\ \bibinfo {author} {\bibfnamefont {J.~M.}\ \bibnamefont
  {Martinis}},\ }\Doi {10.1103/PhysRevA.85.042321} {\bibfield  {journal}
  {\bibinfo  {journal} {Phys. Rev. A},\ }\textbf {\bibinfo {volume} {85}},\
  \bibinfo {pages} {042321} (\bibinfo {year} {2012})}\BibitemShut {NoStop}%
\bibitem [{\citenamefont {Fowler}\ \emph {et~al.}(2011)\citenamefont {Fowler},
  \citenamefont {Wang},\ and\ \citenamefont {Hollenberg}}]{Fowler2011QIC}%
  \BibitemOpen
  \bibfield  {author} {\bibinfo {author} {\bibfnamefont {A.~G.}\ \bibnamefont
  {Fowler}}, \bibinfo {author} {\bibfnamefont {D.~S.}\ \bibnamefont {Wang}}, \
  and\ \bibinfo {author} {\bibfnamefont {L.~C.~L.}\ \bibnamefont
  {Hollenberg}},\ }\href@noop {} {\bibfield  {journal} {\bibinfo  {journal}
  {Quantum Information and Computation},\ }\textbf {\bibinfo {volume} {11}},\
  \bibinfo {pages} {0008} (\bibinfo {year} {2011})}\BibitemShut {NoStop}%
\bibitem [{\citenamefont {Fowler}(2012)}]{Fowl12g}%
  \BibitemOpen
  \bibfield  {author} {\bibinfo {author} {\bibfnamefont {A.~G.}\ \bibnamefont
  {Fowler}},\ }\href@noop {} { (\bibinfo {year} {2012})},\ \Eprint
  {http://arxiv.org/abs/1208.1334} {arXiv:1208.1334 [quant-ph]} \BibitemShut
  {NoStop}%
\bibitem [{\citenamefont {Fowler}\ \emph
  {et~al.}(2012){\natexlab{c}}\citenamefont {Fowler}, \citenamefont
  {Whiteside}, \citenamefont {McInnes},\ and\ \citenamefont
  {Rabbani}}]{2012arXiv1202.6111F}%
  \BibitemOpen
  \bibfield  {author} {\bibinfo {author} {\bibfnamefont {A.~G.}\ \bibnamefont
  {Fowler}}, \bibinfo {author} {\bibfnamefont {A.~C.}\ \bibnamefont
  {Whiteside}}, \bibinfo {author} {\bibfnamefont {A.~L.}\ \bibnamefont
  {McInnes}}, \ and\ \bibinfo {author} {\bibfnamefont {A.}~\bibnamefont
  {Rabbani}},\ }\Doi {10.1103/PhysRevX.2.041003} {\bibfield  {journal}
  {\bibinfo  {journal} {Phys. Rev. X},\ }\textbf {\bibinfo {volume} {2}},\
  \bibinfo {pages} {041003} (\bibinfo {year} {2012}{\natexlab{c}})}\BibitemShut
  {NoStop}%
\end{thebibliography}%

\end{document}